# Harmonic Analysis and Qualitative Uncertainty Principle

Ji King[*]

*Abstract*—This paper investigates the mathematical nature of qualitative uncertainty principle (QUP), which plays an important role in mathematics, physics and engineering fields. Consider a 3-tuple ($K$, $H_1$, $H_2$) that $K: H_1 \rightarrow H_2$ is an integral operator. Suppose a signal $f \in H_1$, $\Omega_1$ and $\Omega_2$ are domains on which $f$, $Kf$ define respectively. Does this signal $f$ vanish if $|\Sigma(f)| < |\Omega_1|$ and $|\Sigma(Kf)| < |\Omega_2|$? The excesses and deficiencies of integral kernel $K(\omega, t)$ are found to be greatly related to this general formulation of QUP. The complete point theory of integral kernel is established to deal with the QUP.

It is shown that the satisfaction of QUP depends on the existence of some complete points and the QUP only holds for good behaved integral operators. An investigation of full violation of QUP shows that $L^2$ space is large for high resolution harmonic analysis. And the invertible linear integral transforms whose kernels are complete in $L^2$ probably lead to the satisfaction of QUP. It indicates the performance limitation of linear integral transforms in harmonic analysis. Two possible ways bypassing uncertainty principle, nonlinear method and sparse representation, are thus suggested. The notion of operator family is developed and is applied to understand remarkable performances of recent sparse representation.

*Index Terms*—Harmonic analysis, uncertainty principle, time-frequency analysis,

---
[*] Ji King is with the School of Engineering, The South China Agricultural University, Guangzhou, Guangdong, 510642, China (e-mail: jiking314@gmail.com).

I. Introduction

Before Gabor's celebrated paper [1] the alternative method of signal analysis had swung between the time and frequency domain as the main apparatus with a very long history [2]. To provide a natural representation of speech comprehension from a signal view Gabor introduced [1] for the first time in communication a new method of analyzing signals which now belongs to the so-called time-frequency analysis where time and frequency play symmetrical parts. This time-frequency method recognizes a signal in two-dimensional space, with time and frequency as coordinates. And the plane is composed of "information diagrams" in terms of both time and frequency conveying the information by a frequency band in a given time interval, their areas proportional to the number of independent information which they can convey. There is the smallest possible area occupied by the certain "elementary signals" conveying one numerical datum each. These elementary signals can readily localize in the time or frequency domain to maximally exploit the data transmission capacity of time interval or frequency band. Nevertheless, the achievable results of these exploitations are restricted by the uncertainty principle. The uncertainty principle states that a nonzero signal or function and its Fourier transform cannot both be sharply localized which is a meta-theorem in harmonic analysis and is a fundamental limitation in many areas, for example, communication [28], control system [151] and even number theory [152]. The time-frequency is such a line merging the advantages of time and

Fourier analysis to adapt to the uncertainty principle. Alternatively the frequency modulation is replaced by the scaling and thus time-frequency becomes time-scale analysis which is usually included in the general framework of time-frequency method.

Over the past sixty years, the time-frequency method has been developed in the mathematical and signal fields [3], [4]. Actually it can be dated back to the original work of Wigner distribution representation of the wave function in quantum physics [5], which was developed as known Wigner-Ville distribution in signal community thereafter [6], [148], [149], [150]. It gives a nearly perfect description of a signal occupying a probability value in the time-frequency region. In quantum theory, points in the time-frequency plane denote states. But the distribution can go negative for some time-frequency points where its physical meaning is thus lost. However, many new time-frequency distributions were proposed to improve Wigner-Ville distribution, see [7], [8], [9], [10], [19] and references therein. Finally these distributions were summarized as the Cohen class [11]. It was proved that there is not a quadratic distribution satisfying the marginal properties and positive condition simultaneously [13]. These distributions fail to assign a probability to each time-frequency points with perfect properties. Besides the time-frequency distributions there are a variety of realizations that localizing signals in time-frequency plane realized. To name a few, there are continuous Gabor transform (also called the short-time Fourier transform or windowed Fourier transform) determining the sinusoidal frequency and phase content of local sections of a signal as it changes over time, frequency modulation splitting the frequency axis into pieces to be frequency-localized (Littlewood-Paley decomposition is such one), wavelet "watching" a signal with different scales obtaining information in detail with an eagle's eye or roughly from a panoramic view, fractional Fourier analysis expanded in a time-angle space.

The above localization strategies in time-frequency plane usually have many different properties. They have their appropriate applications in a wide range of areas. Just to mention some, there are communications [14], signal processing in radar, sonar, geophysical seismology [15], [16], [17], operator approximation [18], functional analysis [29] and quantum mechanics [4]. Actually these approaches are integral transforms which map time signals continuously to other domains through diverse kernel functions. The time signal then has an integral representation in terms of frequency/scale or other variables. Besides integral representations, a signal is also considered as the superposition of elementary functions or basic waves. Time-frequency method were therefore developed in such a way that the signal is described with elementary functions or called fundamental harmonics that are orthogonal or nonorthogonal, exact or redundant, uniform or non-uniform, complete or sparse. In general, the elementary functions are taken as orthonormal basis, Riesz basis, the frame [22], [23], [24] or the atoms [25], [26], [27]. An orthonormal basis provides a unique representation for a signal and the signal is readily reconstructed from its decomposition coefficients. The popular and most studied Fourier series is such an apparatus. The perturbation of Fourier series named nonharmonic Fourier series [23], [30] which is a Riesz basis are then considered, and it is obtainable from a bounded invertible operator acting on Fourier series. Riesz basis constitutes the largest and most tractable class of bases known. By extending Riesz basis, frames were thereafter introduced by Duffin and Schaeffer in the context of nonharmonics [32], which was deliberately developed and was applied to wavelets by Daubechies [22] very late. The potential redundancy of frames often allows them to be more easily constructible than bases, and to possess better properties than achievable bases. It has some advantages, for example, to data transmission with erasures or being robust to noise [34], [35]. The atoms have large redundancies and are more flexible than bases and frames. They serve as basic building blocks for signals in the certain function space, in which any signal is expressed as a sum of these atoms. Here it is especially noted that the atoms stemmed from signal decomposition [25] are somewhat different from, but with more relaxations than the same terminology defined in the standard harmonic analysis [20], [21], [55] and coherent states (also known as atoms in mathematics) [56]. All the collections of atoms are constituted to be a dictionary which can be the unions of pairs of basis or frames. The dictionary is usually to be overcomplete in its analyzed space to give a signal atomic decomposition as a superposition of the least (sparsest) number of atoms according to the sparse representation. It is noticed that basis and frames both do provide complete representations of signals. Recently, however, a sparse approximation and coding was found to be deeply and inherently describing solutions of linear systems of equations or signal models [13], [40]. To be concise, approximating a signal or an image with a sparse linear expansion from a possibly overcomplete dictionary of basis functions

(called atoms) has turned out to be an extremely useful tool to solve many other signal processing problems. Some remarkable and seemingly puzzling results from sparse representation were obtained in the fields of signal recovery [36], [37], compressed sensing [38], sampling theory [39] with performances of a very low sampling rate, extremely high compression ratio and nearly complete signal recovery. Like some truly essential concepts often restated or unearthed in other disciplines, it was also discovered that the strategy of neurons encoding sensory information was referred to as a "sparse" rule, although the definition of "sparse" are slightly different [41] from our stated above. Studies on visual cortex, auditory perception all supported this "sparse" rule [42], [43], [44] and indicated that sparse coding could be a ubiquitous strategy employed in several different modalities across different organisms. Therefore the basis, frames, atoms build the large set of tools in harmonic analysis.

However, in the integral representations or signal decompositions, the uncertainty relation doesn't vanish but have been in form of varieties and extensions of classical uncertainty inequality [27], [45], [46], [47] being an inevitable principle. It is at the heart of harmonic analysis that gives a unified comprehension to different notions in interdisciplines with actual the same mathematical contents, for instance, the microlocal analysis in PDE [48] and time-frequency analysis [49], the Kadison-Singer Problem and the Feichtinger conjecture[*] [50], [51]. Generally the uncertainty principle is recognized imposing the certain limit to resolve functions or spaces in harmonic analysis, also in signal area and other related fields. Unfortunately this principle is so profound that vague ideas occurred in the past resulting in some misleading arguments along with obtained extensive results. However, uncertainty relations are useful devices for estimating signal duration and power spectra in signal analysis. It is well known the spectral density function spreads over infinity for a truncated signal and even for a truncated monotone. The spectrum can disperse widely as the time signal locates within a sharp instant. In this regard usually we say uncertainty principle is working in the Fourier analysis. But if only for the purpose of signal structure identification, the effect caused by uncertainty principle can be actually bypassed in many situations. Now that sparse models of signal allowing an exact signal recovery is a strong support of this viewpoint [36]. The example figuring out this viewpoint most is a monotone identification. It can be identified through the maximum of its spectrum no matter how short this signal is truncated in absence of noise. But when the monotone is contaminated by noise, the exact identification isn't yet achievable. Then again one says it is somehow restricted by uncertainty principle. But it should be noted that the above uncertainty principle affected by noise contamination cannot be confused with other cases wherein the uncertainty principle indeed plays it essential role. In fact the failure to exact identification of noisy signal is associated with the precision of frequency location. And there is a discrepancy in the two terms of precision and resolution. Uncertainty principle works with the latter. Resolution refers to the ability to distinguish two objects, whereas precision is in reference to the accuracy with which a single object can be tracked or identified [44]. The uncertainty principle limits the ability to separate different objects under certain resolution, rather than to track a single signal. Then one should distinguish two cases that an inaccurate signal identification comes from noise contamination, or is subjected to its intrinsic resolution limit and thus to uncertainty principle. Recognizing a signal relies heavily on different representations of a signal. It has been known that the signal representation corresponds to its group structure. The Fourier series is a torus group representation of periodic signals; the locally compact abelian group is associated with the non-periodic signal results in the Fourier integral; the Heisenberg group corresponds to Gabor transform and the affine group leads to wavelet and so on [52], [53], [54]. The accurate harmonic analysis is then quite reduced to identification of group structure of a signal, especially for the truncated one, which seriously muddies the clear group structure embedded in a signal. And speculative attempts to reconstruct a signal from its truncated sample via some certain group representations are remained to be in a dilemma. These considerations lead to the following questions. Whether the uncertainty principle can impose performance limitations on the signal recovery whatever the recovery method or apparatus is employed? Harmonic analysis can go beyond uncertainty principle or not? These questions will be deliberately addressed in this paper.

---

[*] The Kadison-Singer Problem, equivalent to the Feichtinger conjecture [50], is proved to be a stronger version of uncertainty principle [51].

This paper is about an understanding of uncertainty principle. The purpose of the phrase "harmonic analysis" used to title this paper is seeking a sense of equilibrium between the scope of applied harmonic analysis and its applications to the signal area where many issues are also directly named with "harmonic analysis" sometimes. The primary contribution of this paper is to develop a fundamental mathematical quantitative description of the essence of qualitative uncertainty principle in harmonic analysis for a general class of linear integral operators. The principal results of this paper are summarized as follows:

- A. Develop some concepts and obtain results on linear independence of functions to investigate the excesses and deficits of function systems.

- B. Relate the sampling theory to the independence of the function systems. Under some conditions the independence of a function system is equivalent to the existence of sample space which is constructed by this function system.

- C. A new phenomenon called "complete function point" (also simplified as "complete point" sometimes) of integral kernels is discovered being related to the qualitative uncertainty principle. This complete point ensures functions belonging to the integral kernel around this point completeness in its defined space. Its algebraic and geometric properties are well exhibited.

- D. The qualitative uncertainty principles with Fourier transform, Wigner distribution, Gabor transform and wavelet are understood under the framework of complete points.

- E. It is found that qualitative uncertainty principle always holds for an invertible linear integral operator which is complete in $L^2$. If any of above conditions is abandoned, the uncertainty principle is then not satisfied.

- F. When the completeness in $L^2$ of an integral operator is removed, a super-resolution beyond the uncertainty principle is achieved. An operator family is constructed to obtain an exact signal analysis.

In this paper the signal addressed is often restricted in $L^2$ space and linear operators in Hilbert spaces are considered.

The remainder of this paper is organized as follows. In Section II the mathematical models and their hidden physics as generating mechanism of signals are discussed. Section III presents a brief review on the uncertainty principle relevant to the signal. Especially, its physical sense of the concept "frequency" in uncertainty principle is investigated. Section IV is central to this paper. The mathematical preliminary about linear independence of functions is presented. The complete point is developed to address overcompleteness of integral kernels. Its algebraic and geometric properties are exhibited. And some applications are provided. In Section V the qualitative uncertainty principles for Fourier transform, Wigner distribution, Gabor transform and wavelet are proved to be satisfied by identifying complete points of their integral kernels. The relation between uncertainty principle and density of a function sequence is disclosed. The body of the paper continues with Section VI. The satisfaction of QUP mainly depends on the linearity, invertibility and completeness in $L^2$ of integral operators. The uncertainty principle is violated by removing some of three conditions. An operator family exploiting the completeness of integral kernels is suggested to obtain results beyond uncertainty principle. At last Section VII a conclusion is drawn.

## II. Mathematical Models of Signals

The signal from real world indicates dynamics of certain physical quantity which is often assumed to be of the nature of a voltage, current, image, field strength, or other linear ones. They correspond in mathematics to the notion of function or more general distribution that depends on time, on space or on something else. It can be deterministic or statistical, stationary or nonstationary, periodic or non-periodic, continuous or discrete. In signal theory, one is not necessarily interested in the system's specific structure, but rather in the way that the available output signal is transformed. And many methods connected to the transforms were thus proposed to deal with signals. The Fourier series is preferred to periodic signals; but in other cases, it may badly work. In this section it is illustrated that the analysis of signals is deemed depending on its physical generating mechanisms and their respective mathematical models. Therefore the model-based harmonic analysis is suggested. To serve this part, these physical generating mechanisms are distinguished into four cases with respect to different governing differential equations in mathematics: linear differential equation with constant coefficients, wave equation with boundaries, wave equation without boundaries and linear differential equation with variable coefficients. These models are connected with the studies of physical phenomena in diverse scientific fields. For the applications of signals and images we will restrict ourselves to differential equations in terms of one or two variables.

The constant-coefficient linear differential equation considered here is ordinary, which usually comes from harmonic oscillator having close analogs in many fields. The most acquainted harmonic oscillator may be the spring-mass system. Additionally it models phenomena involving a pendulum with a small swing, a RLC electrical circuit, operations of a servo system, complicated interactions in chemical reactions, or the signal transmission in an ideal communication system and so forth [57]. All these phenomena and their generating signals following the equations which are very similar to one another, are subjected to linear differential equations with constant coefficients, and the corresponding system is called the system of a finite number of degrees of freedom [58]. They are the sources of virtually all harmonics and waves. In the following we are going to give an overview of this model to see how the harmonics and associated parametric spectral technique evolving from this linear ordinary differential equation allowing us to treat signals intrinsically produced in such a system. The reader interested in this topic is referred to Courant and Hilbert's famous book [58], from which the outline of our sequel material comes.

Considering a free vibration of the conservative system with $n$ degrees of freedom in the set of independent generalized coordinates $x_1, x_2, \ldots, x_n$, the kinetic energy $T$ and potential energy $U$ of the system are given by the quadratic forms

$$T = \sum_{i,j=1}^{n} a_{ij} \dot{x}_i \dot{x}_j \tag{2.1}$$

and

$$U = \sum_{i,j=1}^{n} b_{ij} x_i x_j \tag{2.2}$$

respectively, with constant coefficients $a_{ij}$, $b_{ij}$ ($1 \leq i, j \leq n$).

From the Lagrange formalism it is known the equations of motion are linear, governed by the differential equations

$$\sum_{j=1}^{n} (a_{ij} \ddot{x}_j + b_{ij} x_j) = 0, \; (i=1, 2, \ldots, n) \; (a_{ij} = a_{ji}, \; b_{ij} = b_{ji}) \tag{2.3}$$

Rewrite the free motion equation with matrices notations

$$AX=0, \qquad (2.4)$$

where $A=[A_1,...,A_n]^T$, $A_i^T =[a_{i1},b_{i1},...,a_{in},b_{in}]$ and $X=[\ddot{x}_1,x_1,...,\ddot{x}_n,x_n]^T$.

Set $Y=BX$, $Y=[\ddot{y}_1,y_1,...,\ddot{y}_n,y_n]$. Let a new coordinate group $\{y_i\}$ ($i$=1, 2, …, $n$) replace the original coordinate $\{x_i\}$ ($i$=1, 2, …, $n$). Substituting $Y$ into (2.4), it has

$$CY=0. \qquad (2.5)$$

Generally $B$ is selected a suitable matrix so as to obtain the canonical form of vibration (2.4). They are

$$\ddot{y}_i + \lambda_i y_i = 0 \ (i=1, 2, ..., n), \qquad (2.6)$$

$\{\lambda_i\}$ are the linear combination of $a_{ij}$ and $b_{ij}$.

It is clear that $\lambda_i$ ($i$=1, 2, …, $n$) are the characteristic values of differential equations (2.4) or (2.5), called eigenfrequenices, or pitches in the vibration. It will be shown the signal generated by a harmonic oscillator system is determined by these generalized coordinates and eigenfrequencies.

A more direct consideration of the above system is by employing the finite dimensional state space. Suppose the states $\{x_i(t)\}$ ($i$=1, 2, …, $n$), then the signal of this system can be taken as

$$x(t)=\sum_{i=1}^{n} \alpha_i x_i(t). \qquad (2.7)$$

The signal $x(t)$ is subjected to the following $n$-order differential equation with constant coefficients

$$x^{(n)} + \beta_{n-1} x^{(n-1)} + ... + \beta_0 x = 0 \qquad (2.8)$$

In the parametric spectral estimation, the identification of constant coefficients $\beta_i$ ($i$=0, 1, …, $n$-1) in (2.8) becomes the main focus, as the signal $x(t)$ is always measured beforehand. The signal from (2.8) is then identified as the superposition of harmonics[*], naturally related to the harmonic analysis. Usually in practice the difference equation is substituted for the differential equation. The AR or ARMA model appears using white noise as an external force driving the difference equations. Then there is no surprise the essence of parametric spectral method coincidences with the linear differential equation of constant coefficients and where the term "parametric" comes from. The formula (2.7) can also be represented by the state space equation, in which it is an output equation and the spectral estimation becomes the modeling of state space.

In the above the wave motion is dominated by the finite eigenfrequencies of the system with discrete medium. But there are also systems having infinity freedoms, or even with continuous medium, for example, the vibration of a stretched string, sound waves in the air, electromagnetic waves or other elastic media. Their governing mathematical model is the wave equation. For the sake of simplicity, it is assumed that a homogenous wave equation transmitted in one dimension (*x*-direction) is the following:

$$u_{xx} = \lambda^2 u_{tt} \qquad (2.9)$$

where $u(x, t)$ is the displacement of the mass from its equilibrium position at a position $x$ and time $t$. The constant $\lambda$ is a parameter related to the speed of propagation of waves in the media.

Suppose the boundary conditions, for example, $u(0, t) = u(\pi, t) = 0$, and the Fourier series are naturally derived from the equation (2.9) by the method of separation of variables and superposition principle [58]. In the electromagnetic theory, the wave equation often originates from Maxell's equations. It well suggests the fact that a great deal of electrical signals is represented by the Fourier series. It is noted that the Fourier series can also stem from heat flow equation, which is the source of discovery of Fourier series.

---

[*] The terms "harmonics", "nonharmonics", "interharmonics" are different from their source definitions but now have been used virtually identical to each other in many situations and the distinctions are not made.

But when begin our study of the wave equation by supposing that there are no boundaries, that is, the wave equation holds for $-\infty < x < \infty$, the eigenvalues of this system form a continuous spectrum clearly. In this case the signal produced from infinite interval wave equation is expanded by the Fourier integral where all the real frequencies replace the denumerable eigenfunction expansion. In signal processing, $L^2(\mathbb{R})$ models signals with finite energies. The Fourier transform is an isometric linear mapping between $L^2(\mathbb{R})$, which provides a way of expanding functions to be a continuous superposition of the basic waves $e^{i\omega x}$ ($\omega \in \mathbb{R}$) as that Fourier series are used to expand periodic functions on a finite interval. $L^2(\mathbb{R})$ is the classical space containing a large class of signals to be considered. Therefore the Fourier transform becomes a standard tool in signal area and studies of spectrum of signals are prevalent with the emergence of FFT. We see that the Fourier integral is also employed to represent finite time functions prescribing their vanishment out of the defined interval. In many cases the applications of nonharmonics, Fourier integral or the Fourier series nearly reach the same results, in particular when only numerical realizations of the three approaches are considered.

The above discussed signals are all stationary regardless of their models from nonharmonics, Fourier series or Fourier integral. The three mathematical tools just model linear time invariant (LTI) systems, which were mostly managed in engineering before the era of time-frequency analysis. However, they are restrictive to be extended to address time-varying signals, for instance, signals in the mobile wireless communications. Transmission over mobile wireless channels undergoing delay spread and Doppler effect will lead to variations of parameters of a system resulting in a time-varying model [59]. The signals of a time-varying system can be described by differential equations with variable coefficients. Generally the inhomogeneous variable coefficient linear differential equation for time-varying systems is

$$u^{(n)}+\alpha_{n-1}(t)u^{(n-1)}+ \ldots +\alpha_{1}(t)u^{(1)}+\alpha_{0}(t)u=f(t) \tag{2.10}$$

The coefficients $\alpha_i(t)$ ($i=0, 1, \ldots, n-1$) are assumed to be continuous in their domains. Suppose the differential operator $L(u)$ associated with left side of (2.10) and solutions of this equation rely on finding a distribution $G(t, x)$ such that $L(G(t, x))=\delta(t-x)$, where $\delta(t-x)$ denotes the Dirac delta function [58]. Formally, then, we can solve $L(u)=f$ by setting

$$u(t)=\int G(t,x)f(x)dx \tag{2.11}$$

if some regular conditions are imposed. This integral kernel called the Green's function $G(t, x)$ usually are not unique. In studies of partial differential equations, (2.11) can be understood within the framework of pseudo-differential operators. It can be realized by wavelet [63], [64] or a superposition of time-frequency shifts of spreading functions having the intrinsic properties of time and frequency localizations [4], [59], [60].

As seen above, harmonic analysis of signals should be based on good identifications of their mathematical models and physical mechanisms. It is not so easy to give an exact idenfification. Firstly the time-frequency implication of (2.11) makes the other three models as specific cases therefore providing universal representations. The nonharmonics from harmonic oscillators can be a Riesz basis or a frame to give a complete representation for finite truncated $L^2$ signals. For any almost periodic function, it is even approximated as the uniform limit of finite superposition of nonharmonics. As to Fourier series, it is often employed to describe a periodic function as well as a function over a finite interval. For the partially observed signal in finite time, the Fourier series provides us a representation when the signal is temporarily regarded as a periodic one. Hence the original structure of this truncated signal and the exact model is uncertain. The Fourier integral can also be an analytic framework for types of signals. To present an example, in optics, the white light contains a mixture of light with different wavelengths colors or different frequencies in equal amounts. It has the decomposition of discrete spectra; nevertheless, a wave train of finite length corresponds not to a sharp line but to a continuous spectrum of finite width which becomes sharper and more intense as the wave train becomes longer and their periodicities can be read from the spectrum. In this case, the finite length wave train is described by a continuous spectrum of Fourier transform of this finite length wave signal. The periodicity of a signal can be identified depending on a cycle data record set obtained or the *priori* knowledge of periodicity that isn't easily affirmed in practice.

For a time-limited signal $f(t)$, let $h(t)$ be the impulse response of a LTI system and $L_h$ be the convolution operator associated with $h(t)$ such that

$$L_h f = (f * h)(t) = \int_{\mathbb{R}} f(t-x)h(x)dx. \tag{2.12}$$

Set $f(t) = e^{i\omega t}$ substituting into (2.12) and it yields

$$L_h e^{i\omega t} = \hat{h}(\omega) e^{i\omega t}, \tag{2.13}$$

where $\hat{h}(\omega)$ is the Fourier transform of $h(t)$, also called the transfer function of this LTI, and $e^{i\omega t}$ is an eigenfunction of the operator $L_h$. One retrieves the extremum value of $\hat{h}(\omega)$ to read the periodicity of $e^{i\omega t}$. It is not a problem for a monotone. But for multi-periodicities signals (superposition of nonharmonics), the search of extremum values is greatly affected by the interference of energy aliasing of $h(t)$. Then the sharp concentration of $\hat{h}(\omega)$ becomes a primary pursuit. The practical signals are often limited spatially or temporally. Lanczos declared the general principle that "a lack of information cannot be remedied by any mathematical trickery" [61]. Clever algorithms are regarded not to produce miracles. Nevertheless, the algorithm with best performance is expected. The model-based harmonic analysis exploits Occam's razor that these models are not equal in goodness of fit accuracy to the observed signals; and there is a best suitable model under some rules, for example, minimum description length principle [62], in that the signal is analyzed. From the uncertainty principle we know there is the certain limit where it fails to resolve very closely spaced signals using Fourier transform. But this failure isn't from the uncertainty principle since the superposition of harmonics free of noise is well estimated via the harmonic oscillator model. In general, the uncertainty relation governs the ultimate limit of spectrum resolution and this estimate error cannot be reduced whatever any clear algorithm is employed. The authors thus conjecture that the *uncertainty principle* in signal area and harmonic analysis borrowed from quantum mechanics isn't the fundamental principle as it is in the quantum world.

## III. Uncertainty Principle

The uncertainty principle, also called "principle of indeterminancy" more descriptively, is one of the fundamental principles in quantum mechanics and has profound implications for such fundamental notions like causality and predictions. It can have so many implications and unspeakable ones remained to be known, which now are still arguable [65], [66], [67], [68], although it had been laid as the foundation for what became known as the Copenhagen interpretation of quantum mechanics applied in a wide variety of fields. In this section we don't want to be plagued into the foundation of quantum mechanics and its ontological or epistemological aspects too much, but more is restricted to its implications and applications to signals. The reader interested in its mathematical development is referred to the good tutorial by Folland and Sitaram [69] where various mathematical uncertainty relations were comprehensively discussed and to the treatise by Havin and Jöricke [127] for advanced results, also is referred to [70], [71] for its physical respects.

The uncertainty principle has versatile significances. It is a description of a characteristic feature of quantum mechanical systems about the nature of the system itself as described by equations of quantum mechanics. It also relates to the observer effect which is a statement about the limit of one's ability to perform measurements on a system without disturbing it. Thus it refers to errors of measurement and to the spread of values of the physical variables intrinsic to a particle's state both [72]. There is a succinct statement of original uncertainty relation that the more precisely the position is determined, the less precisely the momentum is known in this instant, and vice versa, which was considered by Heisenberg first in 1927. In his seminal paper [73] the Heisenberg uncertainty relation was derived as a thought experimental consequence of a measuring process for resolving power of a gamma-ray microscope. For every measurement of the position $x$ of a mass with root-mean-square error $\Delta x$, the root-mean-square disturbance $\Delta p$ of the moment $p$ of the mass caused by the interaction of this measurement always satisfies the relation $\Delta x \Delta p \geq \hbar/2$, which was reformed by Kennard [74] immediately as the famous inequality for the standard deviations of position and momentum. Actually Weyl [75] also achieved the similar mathematical results. Robertson [76] in 1929 generalized Kennard's relation to noncommutative observables. It states that for any pair of observables (self-adjoint operators) $A$ and $B$, it has

$$\Delta_\psi A \Delta_\psi B \geq |<\psi, [A, B]\psi>|/2 \qquad (3.1)$$

In above, $\psi$ denotes any state of a particle, $[A, B]=AB-BA$ stands for the commutator, and $\Delta_\psi A$, $\Delta_\psi B$ are the standard deviations of observables $A$, $B$ in the system state $\psi$ respectively. The uncertainty principle was thus laid on the rigorous mathematical foundation.

The uncertainty principle appears us in a general harmony form, as one always sees, the meta-theorem claiming that it is not possible for a non-trivial function and its Fourier transform to be simultaneously sharply concentrated, which gains a universe validity in harmonic analysis and mathematical aspects of quantum theory in this abstract elaboration. Its quantitative version says that for each pair of conjugate variables, the product of the uncertainties of both can never be smaller than a fixed value specified by respective uncertainty relations. To present a precision form, it has

$$(\Delta_\alpha f)(\Delta_\beta \hat{f}) \geq 1/4 \qquad (3.2)$$

for any $\alpha, \beta \in \mathbb{R}$. Here there are

$$\Delta_\alpha f = \int_\mathbb{R} (t-\alpha)^2 |f(t)|^2 \, dt \Big/ \int_\mathbb{R} |f(t)|^2 \, dt, \quad \Delta_\beta \hat{f} = \int_\mathbb{R} (\omega-\beta)^2 |\hat{f}(\omega)|^2 \, d\omega \Big/ \int_\mathbb{R} |\hat{f}(\omega)|^2 \, d\omega.$$

The Fourier transformation of $f$ is defined as

$$\hat{f}(\omega) = \int_\mathbb{R} f(t) e^{-i\omega t} \, dt.$$

The notations $\Delta_\alpha f$ and $\Delta_\beta \hat{f}$ are the measures of how much $f$ and $\hat{f}$ fails to be localized near $\alpha$ and $\beta$. If $\alpha$ is the expectation of $f$, then $\Delta_\alpha f$ is the variance being the dispersion of $f$ about the point $\alpha$. The inequality (3.2) is also called Heisenberg inequality to be widely used. The Heisenberg uncertainty inequality has great effects on related topics of signals. It in a broad sense tells us how much the frequency contents which make the reconstruction of original functions possible. It is known intuitively from (3.2) that enough harmonic components are needed to produce the time concentrated signal or vice verse. This required total amount of harmonics is referred to as spectral bandwidth. Essentially, a spectral bandwidth is a standard measure of spectral dispersion of signal power in terms of frequencies. Beyond the time-frequency uncertainty relations, there is a extended term "conjugate variables" referring to a pair of variables that they becomes Fourier transform dual one-another, or more generally referring to the Pontryagin duals. The uncertainty principle is always well derived from conjugate variables. In quantum mechanics, there is an associated operator corresponding to each physically observable quantity. Therefore the noncommutative operators stand for conjugate variables. The prominent example that position measure with $A\psi=x\psi$, and its momentum with $B\psi=d\psi/dt$ is the special case of (3.1) governed by $[A, B]=-I$ (the identity operator). However, in actual fact their noncommutativity of two observables or operators isn't certain to ensure the existence of uncertainty principle, which had been discussed in [69] and [77].

    The uncertainty principle is of importance in the signal analysis taking the time and frequency these two essential notions in the signal field as conjugate variables. But it was perhaps the first person Bohr in 1928 that had realized such a kind of uncertainty relations from classical wave theory known as the bandwidth theorem [78], [79]. The Bohr's uncertainty relation concerns waves of finite length, i.e., wave pulses, and states wave pulses do not possess a single precise frequency or wavelength but rather consist of a range of them. The bandwidth relation is based on the representation of waves of finite length by Fourier series, according to which any periodic function is equivalent to a sum of infinite sine waves of different frequencies and wavelengths. However, the first explicit introduction of the exact formulation of uncertainty relation to signal community should be credited to Gabor [1]. His work appeared that

time to be unique in this area following in an unusual direction introducing quantum effect into the derivation of information rates and a consideration for channel capacity while not analyzing quantum systems with information theory. With a communication channel, Gabor considered the description of a signal in terms of time and frequency both, which contribute in quantum information diagram conveying exactly one datum. There is a minimal information diagram, that is, optimal time-frequency localized elementary function, acquired through exact formulation of uncertainty relation. Since then, this uncertainty principle "has become firmly embedded in the common culture" [69]. Thereafter the uncertainty relation is treated under more practical conditions with various concrete manifestations. These specific quantitative relations strongly depend upon interpretations of "concentration" which can be always taken as the uncertainty principle in disguise. For example, the uncertainty inequality is different for a real signal from the inequality for a complex signal [80]; the uncertainty relation also changes when non-casual signal is replaced by a one-sided casual signal [81]. But when the measure of "concentration" that ratio of signal's energy in a piece of time duration or frequency band to its total energy is adopted, there were elegant results contributed by Landau, Pollack and Slepian in their serial papers [82-86]. Considering a time and band-limited signal, there is a function occupying a maximal energy concentration in the time domain given a frequency band and vice verse. Such a function being satisfied with optimal properties is called the prolate spheroidal wave function (PSWF). Note that the PSWF doesn't have the super resolving power for two frequency closed signals, as the spectrum resolution is also set a limit by its mainlobe width of PSWF. However, the concept of time and band-limited functions was further exploited in sequential papers in diverse areas [87-91], [96], [97] and simultaneously the PSWFs were investigated in a board sense and with an in-depth insight [92-95]. Now the PSWFs seem likely to become increasingly popular in communication and related fields [98], [99].

From above discussion the uncertainty principle in signal analysis may be primarily considered only in a formal sense not as it is in quantum mechanics. But optic communication, bridging the gap between this substance of uncertainty principle in quantum world and the formalism in signal theory, shows us the more implications of uncertainty principle for signals. When the transmit rate of a communication system runs with a high frequency, quantum effects become important where the uncertainty relation derived from the point of view of signal theory contains the same physical significance as this principle from quantum mechanics. This case can be understood from the uncertainty principle in special relativity. It is clear the energy-time uncertainty relation $\Delta t \Delta E \geq \hbar/2$ holds since energy bears the same relation to time as momentum does to space in special relativity. The quantity time refers to the moment when the photon is measured arriving at. And the energy is defined by its frequency based on the de Broglie formula $E=hf$. Here the frequency definition in this formula is absolutely unambiguous. But we note that it is the frequency that wild applications of this arguable notion make the room for various uncertainty relations in harmonic analysis.

When recall the basic definition of frequency in physics, it originally refers to the number of oscillations per unit time from a single periodic vibratory motion. This source definition is related to the time, also called temporal frequency. However, the frequency (temporal frequency) akin to its spatial analog of wavenumber (spatial frequency), its exact definition varies in different research areas. Similar to the local wavenumber, there are also terminologies applied such as the "changing frequency" or "instantaneous frequency" to describing aperiodic signals. These terms contained vague senses and great efforts were made in order to clear them [101-104]. But there is not a general definition of frequency consistent with all cases. Each had its suitable situations. In an application, no seeming paradoxes could be made when one agrees an explicit meaning of frequency. The status quo of murky definitions for the most part comes from the Fourier frequency of signals which is not a physical reality. Einstein et al in their famous EPR paper [105] posed one such a way of recognizing a physical reality in agreement with classical as well as quantum mechanism that "if, without in any way disturbing a system, we can predict with certainty (i.e., with probability equal to unity) the value of a physical reality corresponding to this physical quantity". Although it is believed to be merely served as a sufficient, not a necessary condition of reality, in most cases it is also the criterion. Supposed a signal $\psi(t)$ ($t<0$) before the origin were known by a measurement apparatus, one can then evaluate with the help of Fourier transform the Fourier frequency contents $\hat{\psi}(\omega)$ for $t<0$. But the future frequency contents of $\psi(t)$ for $t\geq0$ cannot be predicted with probability unity, and the obtained data set $\psi(t)$ ($t\geq0$) may collapse the old frequency pattern $\hat{\psi}(\omega)$ for $t<0$ instead a new pattern $\hat{\psi}_1(\omega)$, Fourier transform of $\psi(t)$ adding new information for $t\geq0$. It indicates that the Fourier frequency cannot be obtained from the measured signal but is highly relevant to the future data which hasn't be measured yet. It is so inconsistent! It is for this contradiction Gabor introduced time-frequency analysis. And the physical quantity "Fourier frequency" is just maintained for its computational convenience and formal analysis while not having real meanings of a physical reality. This non-casual ambiguous concept calls for a reasonable interpretation and a clear idea of frequencies of a signal.

Gabor discussed there was an alternative depiction of the frequency replacing sine waves, which was satisfied a system that was built up with most simplicity, for possible orthogonal functions employed in communications acting as frequencies [1]. On addressing uncertainty principle, there was the eigenfunction substitute for Fourier frequency to achieve the uncertainty relation in the case of eigenfunction expansions [107]. In quantum mechanics there are eigenfunctions corresponding to the physical quantities measured simultaneously, i.e. all have definitive values at the same time independent of the measurement. All such sets of physical quantities which give a complete description of a state in quantum mechanics constitute known as complete sets [106]. If we consider the "new frequency" represented by commutative operators instead of Fourier transform, then it would have the definitive value both at time and frequency domains compatibly. In the past many papers actually dealt with the uncertainty principle with new frequency from the point view of group representations [108-111]. Hogan [108] discovered that the satisfaction of the qualitative uncertainty principle for a particular locally compact group $G$ was largely determined by the topology of *G* and argued that if *G* be noncompact, nondiscrete Plancherel group with a compact open normal subgroup $G_1$. Then the qualitative uncertainty principle (QUP) can be violated by *G*. It is suggested the QUP was sharply dependent on the topology of original group thus its transformation (group representation). The preceding observations motive the author to exhaustively examine the conditions on which the uncertainty principle is satisfied and the nature of how uncertainty principle works in nature. Such a problem has been remained open to the author's knowledge, but now may be attacked afresh in some other ways.

IV. Linear Independence, Integral Kernels and Complete Points

This section is the main body of the paper, which is central to uncertainty principle and its applications. It investigates linear independence, excesses and deficits of function sequences. To extend these notions to continuous index function systems, a new phenomenon "complete point" theory is explicitly developed. Its fundamental properties, such as existence, algebraic and geometric properties are presented. Some special kernels including Hilbert-Schmidt kernels, reproducing kernels, translation invariant kernels are addressed within the framework of complete point theory. Here we will start from a general uncertainty principle statement associated with an integral operator.

Let *K* be an integral operator that $K: H_1 \to H_2$, and $H_1$, $H_2$ are Hilbert function spaces in which the functions are defined on two domains $\Omega_1, \Omega_2 \subseteq \mathbb{R}$ respectively. Now the general qualitative uncertainty principle (QUP) associated with the 3-tuple ($K$, $H_1$, $H_2$),

similar to the Fourier case, says as the following:

Suppose $f \in H_1$. if $|\Sigma(f)| < |\Omega_1|$ and $|\Sigma(\tilde{f})| < |\Omega_2|$, then $f=0$.  (4.1)

where $\tilde{f}$ is the transformed function of $f$ under the operator $K$, two set $\Sigma(f)=\{x: f(x) \neq 0\}$, $\Sigma(\tilde{f})=\{\omega: \tilde{f}(\omega) \neq 0\}$ and $|\cdot|$ denotes the Lebesgue measure of a set.

Some results about group representation similar to this assertion (4.1) were assembled in [69]. This QUP is to be investigated with a general class of 3-tuple ($K$, $H_1$, $H_2$) to examine conditions under which QUP is preserved or not. It will be shown that global properties of integral operators are highly related to the QUP. Three parts of this section are served to analyze integral operators and the QUP. The first part is about the linear independence of function systems, which now is of importance in the research of frame theory [112]. Recently some papers [113-121] have studied density, overcompleteness and localizations of frames. Not restricted to frames and function sequences that are called discrete index function system in this paper, the notion of linear independence is extended to the vector-valued functions also referred to as continuous index function system, and the excesses and deficits of function systems are defined based on it. In the second part the local and global analysis of integral kernels with the concept of "complete points" are presented, as well with its algebraic and geometric properties of the complete point. In the third part (simple) complete points are extended to assembly complete function point set.

*A. Linear Independence of Function Systems*

In this subsection the linear independence of function systems is investigated. First some necessary definitions are given below.

Definition 4.1. A function system is a function set $\{K_\omega(t)|K_\omega(t)\in H, t\in I, \omega\in J\}$, $I, J \subseteq \mathbb{R}$, $I$ an interval, $H$ the function space. If $J$ is a discrete index set, $\{K_\omega(t)\}$ is called a discrete index function system referring to the usual function sequences, or simplified as a discrete function system (DFS). If $J$ is a continuous index set, $\{K_\omega(t)\}$ is a continuous function system (CFS).

It is noted here a continuous index set can be a single interval or a union of intervals. And usually when we say the dimension of function systems, it actually refers to the dimension of their spanned space. But if the spanned space is infinite, it is then difficult to measure the excesses and deficits of function systems, especially for a CFS, although its spanned space may be the same as that spanned by a DFS.

Definition 4.2. Let $\{K(\omega,t)\}_{(\omega,t)\in J\times I}$ be a function system, $I, J \subseteq \mathbb{R}$, $K_\omega(t)\in L^2(I)$ in terms of $t$ for any $\omega\in J$, and the space $H$ spanned by this function system. Now a subsystem $\{K(\omega,t)\}_{(\omega,t)\in \Omega\times I}$, $\Omega \subseteq J$, with its spanned space $H_1$, is said to be

(i) $\Omega$-complete, if $H_1=H$.

(ii) $\Omega$-incomplete, if $H_1 \subset H$.

(iii) $\Omega$-overcomplete, if there exists $\Omega_1 \subset \Omega$ such that $\overline{\text{span}}\{K(\omega,t)\}_{(\omega,t)\in \Omega_1\times I}=H$.

According to this definition, an $\Omega$-overcomplete function system is also $\Omega$-complete. This agreement will work throughout this paper. The terminology "span" refers to the "finite linear span" which is the set of all finite linear combinations of elements of $\{K_i(t)\}$. A space $H$ spanned by $\{K_i(t)\}$ is exactly the closure of finite linear span of $\{K_i(t)\}$. $\{K_i(t)\}$ is said to be complete in the space $H$ if span$\{K_i(t)\}$ is dense in $H$, or $<f, K_i(t)>=0$ for all functions $\{K_i(t)\}$ leading to $f=0$ [23]. And the space spanned by a continuous function system is defined to be the closed linear span of $\{K(\omega_i, t)\}$ for any distinct point $\omega_i \in J$. There are some intuitional results concerning completeness of DFS by analogy to the vectors in linear spaces as follows.

Example 4.3. Let $\{K(\omega,t)\}_{(\omega,t) \in J \times I}$ be a DFS. If the linear independence holds for this system, then it is $\Omega$-incomplete when $\Omega \subset J$. If it is linearly dependent, then there is a set $\Omega \subset J$ such that it is $\Omega$-complete.

The finite dimensional DFS has been well developed in mathematics and there are some tools applicable to address its linear independence. The Wronskian determinant is such an apparatus allowing one to decide its linear independence directly.

Definition 4.4. By the Wronskian matrix of a system of numerical functions $K_i(t): I \to \mathbb{R}$, $i=1, \ldots, n$ ($I$ an interval) is meant the numerical matrix $W(K_1,\ldots, K_n)(t)$ whose element at $(i, j)$ equals

$$W(K_1,\ldots, K_n)(t)_{i,j} = K_j^{(i-1)}(t) . \qquad (4.2)$$

We denote the Wronskian determinant $|W(K_1,\ldots, K_n)(t)|$ or $\det(W(K_1,\ldots, K_n)(t))$.

The Wronskian matrix requires the functions to be differentiable enough. But the vast majority of signals addressed belong to $L^2$. Then the general Wronskian matrix is developed.

Definition 4.5. Suppose the numerical functions $K_i(t): I \to \mathbb{R}$, $i=1, \ldots, n$ ($I$ an interval), $K_i(t) \in L^2$ and $h(t)$ a mollifier. It calls the general Wronskian matrix as it is defined

$$W_p(K_1,\ldots, K_n)(t)_{i,j} = K_j(t) * h^{(i-1)}(t), \qquad (4.3)$$

where the symbol * denotes convolution and $h^{(i-1)}(t)$ is the ($i$-1)-times differentiation of $h(t)$.

The mollifier $h(t)$ is a function belongs to $C^\infty$ (smooth function space) to ensure the analytic property of entries of $W_p(t)$ [122].

Theorem 4.6. Let $K_i(t)$ ($i=1,\ldots,n$) be $C^\infty$ (or $L^2$) numerical functions on the interval $I$, and $W(K_1,\ldots, K_n)(t)$ (or $W_p(K_1,\ldots, K_n)(t)$) the (general) Wronskian matrix constructed by these functions. Then there are

(i) If $|W(t)|\neq 0$ ($|W_p(t)|\neq 0$) at some points of $I$, then $K_1,\ldots, K_n$ are linearly independent on $I$.

(ii) If $|W(t)|\equiv 0$ ($|W_p(t)|\equiv 0$) over $I$, and $K_1,\ldots, K_n$ are analytic, then they are linearly dependent on $I$.

Proof: For the Wronskian matrix $W(t)$, (i) and (ii) are standard results in the mathematics [123]. Only the general case $W_p(t)$ is proved here.

Suppose there is a vector $\boldsymbol{a}^T=(a_1, a_2,\ldots, a_n)$ such that $\sum_{j=1}^{n} a_j K_j(t)=0$. Then there are

$$(\sum_{j=1}^{n} a_j K_j(t))* h^{(i-1)}(t)=0. \quad (i=1,\ldots, n) \tag{4.4}$$

That is, $\qquad W_p(t)\boldsymbol{a}=0. \tag{4.5}$

If $|W_p(t)|\neq 0$ at some points of $I$, the solution vector of (4.5) is zero. Therefore $K_1,\ldots, K_n$ are linearly independent on $I$. But if $|W_p(t)|$ vanishes over the interval $I$ and $K_j*h(t)$ ($i=1,\ldots, n$) are analytic, then there exists nontrivial solution vector $\boldsymbol{a}$ admitting (4.5) from (ii) of Wronskian matrix $W(t)$ [123]. There is $K_1*h(t),\ldots, K_n*h(t)$ linear dependence on $I$. Hence $K_1,\ldots, K_n$ are linearly dependent on $I$. \hfill END

It sees that their linear independence of these functions can also be determined by the Gram determinant [58]. But the sampling theory presents a more general and applicable framework to analyze linear independence of function sequences not confined to smooth functions. Concerning the finite dimensional DFS, there are simple sampling formulae. In the finite dimensional sampling theory, there is the corresponding sampling matrix to the sampling points $\{t_1, t_2,..., t_n\}$ defined by

$$\Psi = \begin{bmatrix} K_1(t_1) & K_1(t_2) & . & . & . & K_1(t_n) \\ . & & & & & . \\ . & & & & & . \\ . & & & & & . \\ K_n(t_1) & K_n(t_2) & . & . & . & K_n(t_n) \end{bmatrix}.$$

Proposition 4.7. For a finite DFS $\{K_i(t)\}_{i=1}^n$ in $L^2$ space, the sufficient and necessary condition of this DFS to be linearly independent is that there are a full rank sampling matrix $\Psi$ and a linearly independent sequence $\{S_1(t), S_2(t),..., S_n(t)\}$ such that $K_i(t) = \sum_{j=1}^n K_i(t_j) S_j(t)$ $(1 \leq i \leq n)$.

Proof: The sufficiency of this proposition is readily to be obtained. We are to prove its necessary part. The following argument is to be proved firstly.

There always exist points $\{t_1, t_2,..., t_n\}$ such that $\Psi$ has full rank.

When $n=1$, the rank of $\Psi$ is full if $K_1(t_1) \neq 0$ at a point $t_1$ is selected. Suppose that $n=N$, the rank of $\Psi$ can also be full for the sampling set $\{t_1, t_2,..., t_N\}$. Therefore for the case of $n=N+1$ there is a new sampling points $t_{N+1}$ such that the rank of new sampling matrix

$$\begin{bmatrix} K_1(t_1) & . & . & . & K_1(t_N) & K_1(t_{N+1}) \\ . & & & & . & . \\ . & & & & . & . \\ . & & & & . & . \\ K_N(t_1) & . & . & . & K_N(t_N) & K_N(t_{N+1}) \\ K_{N+1}(t_1) & . & . & . & K_{N+1}(t_N) & K_{N+1}(t_{N+1}) \end{bmatrix} \quad (t_{N+1} \neq t_1, t_2,..., t_N)$$

also be full. Otherwise it will have and only have a nonzero basic solution vector $a^T=$ $(a_1, a_2,\ldots, a_{N+1})$ satisfying

$$\begin{bmatrix} K_1(t_1) & . & . & . & K_N(t_1) & K_{N+1}(t_1) \\ . & & & & . & . \\ . & & & & . & . \\ . & & & & . & . \\ K_1(t_N) & . & . & . & K_N(t_N) & K_{N+1}(t_N) \\ K_1(t_{N+1}) & . & . & . & K_N(t_{N+1}) & K_{N+1}(t_{N+1}) \end{bmatrix} \begin{bmatrix} a_1 \\ . \\ . \\ . \\ a_N \\ a_{N+1} \end{bmatrix} = 0 \qquad (4.6)$$

for any $t_{N+1} \neq t_1, t_2,\ldots, t_N$. It means that $\sum_{i=1}^{N+1} a_i K_i(t) = 0$ for any $t$ from (4.6). But $\{K_1(t), K_2(t),\ldots, K_n(t), K_{N+1}(t)\}$ are linearly independent according to the given conditions of this proposition and our supposition. Then this is a contradiction, therefore the full rank sampling matrix $\Psi$ exists. To solve the equation

$\Psi[S_1,\ldots, S_n]^T = [K_1,\ldots, K_n]^T$.

It has

$[S_1,\ldots, S_n]^T = \Psi^{-1}[K_1,\ldots, K_n]^T$, which is linearly independent.   END

Proposition 4.8. For a finite DFS $\{K_i(t)\}_{i=1}^n$ in $L^2$ space with the rank $m$, then there are a point set $\{t_i\}_{i=1}^m$ and independent functions $\{S_i(t)\}_{i=1}^m$ such that $K_i(t) = \sum_{j=1}^m K_i(t_j) S_j(t)$ $(1 \leq i \leq n)$.

Proof: With the rank $m$, there are $m$ functions, supposing $\{K_i(t)\}_{i=1}^m$ without loss of generality, of linear independence. From Proposition 4.7, it has

$$K_i(t) = \sum_{j=1}^m K_i(t_j) S_j(t) \quad (1 \leq i \leq m). \qquad (4.7)$$

Then for $m < p \leq n$, it has $K_p(t) = \sum_{i=1}^m a_{pi} K_i(t)$

$$= \sum_{i=1}^{m} a_{pi} (\sum_{j=1}^{m} K_i(t_j) S_j(t))$$

$$= \sum_{j=1}^{m} (\sum_{i=1}^{m} a_{pi} K_i(t_j)) S_j(t)$$

$$= \sum_{j=1}^{m} K_p(t_j) S_j(t). \tag{4.8}$$

Combining (4.7) and (4.8), the proposition is proved. END

To extend the finite function system to its infinite case, an $\omega$-independence is adopted if the series $\sum_{i=1}^{\infty} a_i K_i(t)$ converges and equals the zero vector only when every $a_i=0$. The infinite dimensional sampling shows very different results from its finite condition.

Proposition 4.9. Let $\{K_i(t)\}_{i=1}^{\infty}$ be a basis in $L^2(I)$ ($I \subseteq \mathbb{R}$). Then there are no sampling points $\{t_i\}_{i=1}^{\infty}$ and independent functions $\{S_i(t)\}_{i=1}^{\infty}$ such that $K_i(t) = \sum_{j=1}^{\infty} K_i(t_j) S_j(t)$ for any $i$.

Proof: It is clear a subset $J \subset I$ is achievable to make sampling points $t_i \in J$. For any function $f \in L^2(I)$ vanishing on $J$, there exists a nontrivial sequence $\{a_i\} \in l^2$ such that $f(t) = \sum_{i=1}^{\infty} a_i K_i(t)$ and $\{f(t_j)\}_{j=1}^{\infty}$ are all zeros. Suppose this proposition is violated. Then it has

$$f(t) = \sum_{i=1}^{\infty} a_i K_i(t) = \sum_{i=1}^{\infty} a_i (\sum_{j=1}^{\infty} K_i(t_j) S_j(t)) = \sum_{j=1}^{\infty} (\sum_{i=1}^{\infty} a_i K_i(t_j)) S_j(t) = \sum_{j=1}^{\infty} f(t_j) S_j(t) = 0, \forall t \in I$$

But it is impossible since $\{K_i(t)\}_{i=1}^{\infty}$ is a basis and $f(t)$ can be set just vanishing on $J$, not on the whole set $I$. END

As a consequence, there is a direct corollary.

Corollary 4.10. $L^2$ cannot be sampled.

It is noted that the precise definition of sampling hasn't been given, but it has been shown following the way in the proposition 4.7-4.9. The linear independence of infinite dimensional function system isn't characterized simply by the sampling theory. As to the finite dimensional CFS we attempt to give a definition of its linear independence.

Definition 4.11. Let $\{K_\omega(t)|(\omega, t) \in J \times I\}$ be a finite dimensional CFS in $L^2(I)$. The rank of this system is $n$ means that there are $\{\omega_i\}_{i=1}^{n} \in J$ such that $\{K_{\omega_i}(t)\}_{i=1}^{n}$ are linearly independent, but is dependent for any $(n+1)$ number functions $\{K_{\omega_i}(t)\}_{i=1}^{n+1}$.

Similar to the DFS, the Wronskian matrix also can be applied to determine the linear independence of some types of CFSs. Given a smooth CFS $\{K_\omega(t)\}$, its $n$-order Wronskian matrix is denoted by

$$W_n(\omega, t) = \begin{bmatrix} K(\omega,t) & \frac{\partial K(\omega,t)}{\partial \omega} & \cdots & \frac{\partial^{n-1} K(\omega,t)}{\partial \omega^{n-1}} \\ \frac{\partial K(\omega,t)}{\partial t} & \frac{\partial^2 K(\omega,t)}{\partial t \partial \omega} & \cdots & \frac{\partial^n K(\omega,t)}{\partial t \partial \omega^{n-1}} \\ \cdot & \cdot & \cdot & \cdot \\ \cdot & \cdot & \cdot & \cdot \\ \cdot & \cdot & \cdot & \cdot \\ \frac{\partial^{n-1} K(\omega,t)}{\partial t^{n-1}} & \frac{\partial^n K(\omega,t)}{\partial t^{n-1} \partial \omega} & \cdots & \frac{\partial^{2n-2} K(\omega,t)}{\partial t^{n-1} \partial \omega^{n-1}} \end{bmatrix}$$

if the partial derivatives exist and are continuous.

Theorem 4.12. Let $\{K_\omega(t)|(\omega, t) \in J \times I\}$ be a continuous function system. If its Wronskian determinant $|W_n(\omega, t)|$ exists and doesn't vanish on $J \times I$, then the rank of $\{K_\omega(t)\}$ is at least $n$ and vice verse.

Proof: First part: If the rank of $\{K_\omega(t)\}$ is at least $n$, then $|W_n(\omega, t)|$ doesn't constantly vanish on $J \times I$.

Assume a nonzero vector $\boldsymbol{a}^T = (a_1, a_2, \ldots, a_n)$ such that

$$W_n^T(\omega, t)\boldsymbol{a} = 0. \tag{4.9}$$

Then there is

$$a_n \frac{\partial^{(n-1)} K(\omega, t)}{\partial t^{(n-1)}} + \ldots + a_1 K(\omega, t) = 0. \tag{4.10}$$

For the differential equation

$$a_n y^{(n-1)} + a_{n-1} y^{(n-2)} + \ldots + a_2 y^{(1)} + a_1 y = 0, \tag{4.11}$$

its solution is $y = K(\omega, t)$. Because the dimension of the solution space of (4.11) is at most $n-1$, and so the rank of $\{K(\omega, t)\}$ is also at most $n-1$. Therefore, if the rank of $\{K(\omega, t)\}$ is greater than $n-1$, i.e. at least $n$, then the solution of (4.9) $\boldsymbol{a}$ must be a zero vector. So it indicates $|W_n(\omega, t)|$ doesn't vanish on $J \times I$.

Second part: If $|W_n(\omega, t)|$ doesn't vanish on $J \times I$, then the rank of $\{K_\omega(t)\}$ is at least $n$. Set $\boldsymbol{K}(\omega, t) = (K(\omega, t), \frac{\partial K(\omega, t)}{\partial t}, \ldots, \frac{\partial^{n-1} K(\omega, t)}{\partial t^{n-1}})$. There are at least $n$ functions $\{\boldsymbol{K}(\omega_1, t), \ldots, \boldsymbol{K}(\omega_n, t)\}$ linear independence. Otherwise the vector-valued function $\boldsymbol{K}(\omega, t)$ can be expanded by at most $n-1$ functions. There is

$$\boldsymbol{K}(\omega, t) = \sum_{i=1}^{n-1} \boldsymbol{K}(\omega_i, t) S_i(\omega).$$

However, it leads to the linear dependence of the functions $\{\boldsymbol{K}(\omega, t), \partial \boldsymbol{K}(\omega, t)/\partial \omega, \ldots, \partial^{n-1} \boldsymbol{K}(\omega, t)/\partial \omega^{n-1}\}$ if we differentiate both sides of the above expansion $i$ times ($i=0, \ldots, n-1$) and make an easy verification. That is, the column vectors of $W_n(\omega, t)$ are linearly dependent. But it is violated by the fact of nonvanishment of $|W_n(\omega, t)|$. Therefore $\{\boldsymbol{K}(\omega_1, t), \ldots, \boldsymbol{K}(\omega_n, t)\}$ is linearly independent and so is the function

set$\{K(\omega_1, t),\ldots, K(\omega_n, t)\}$. According to definition 4.11 then there is Rank$\{K(\omega, t)\}\geq n$ when $|W_n(\omega, t)|$ exists and doesn't vanish on $J \times I$.                                    END

In the infinite dimensional system, the "independence" doesn't ensure the meaning of "spanning" where there are many "shades of gray" as argued by Heil in [112]. Heil called it a *Property S* that any function of its space can be expanded by a series. But if a function sequence is linearly dependent, then any function of them is expanded by this sequence. Therefore it automatically has the *Property S*, which is a useful result to be frequently applied in this paper. For more quick introductory materials about this, the reader is referred to Heil's elegant manuscript [124].

As the infinite dimensional and continuous functions are difficult to be addressed directly, then we return to Definition 4.2. The definition is extended to be applied in the next part.

*B. Integral Kernels and Complete Points*

In this subsection the foundation, called complete function point (CFP) theory, underlying the qualitative uncertainty principle is disclosed, as well as its properties exhibited. It shows that the satisfaction of QUP depends on the density of integral kernel and the study of CFPs of integral kernels will lead to the useful investigation of some special kernels. The subsection contains three parts. The first part introduces the CFP phenomenon and some its properties for a general class of integral operators. In the second part the geometric theory of CFP is established. Applications to some special integral kernels are studied in the third part.

*B1. Complete Function Points and Their Properties*

First we define an integral operator on the Hilbert space. Some constraints will be imposed on the integral operator for needed.

Definition 4.13. Let $(I, \Sigma, \mu)$ be a $\sigma$-finite measure space with $\mu$ positive and a measurable function $f$ on $I$. A linear map $K$ with domain $H_1$ and range $H_2$ is called an integral operator if there is an integral kernel, which is denoted by $K(\omega, t)$ defined on $J \times I \subseteq \mathbb{R}^2$ such that

$$\tilde{f}(\omega) = K(f)(\omega) = \int_I K(\omega, t) f(t) d\mu(t), \text{ for any } f \in H_1. \qquad (4.12)$$

In most situations the simple case $d\mu(t)=dt$ endowed with Lebesgue measure on the real axis is employed. In this paper $H_1$ and $H_2$ are supposed to be Hilbert spaces and often belong to $L^2(\mathbb{R})$ unless otherwise stated. This integral operator has diverse properties provided its kernel $K(\omega, t)$ with different conditions. Usually the smooth integral operator is addressed. We don't deal with the singular integral operator, which is outside the scope of this paper. The operator should be assumed to be bounded and invertible. There are conditions which provide such a general class of integral operators, called continuous frames if there are positive values $A$, $B$ such that $A\|f\|^2 \leq \|\tilde{f}(\omega)\|^2 \leq B\|f\|^2$ for every $f \in H_1$. In the special case that $A=B$, the continuous frame is called tight, for which the Plancherel-like formula is preserved [125], [126]. It is clear that popular integral transforms, for example, Fourier transform, wavelet are included in this setting of continuous frames. When $J$ is a discrete index set, $\{K(\omega, t)\}$ would become a discrete frame system.

*General Definitions and Properties*

Since the uncertainty principle indicates nonvanishments of functions and their transforms both within some regions, it is essential to investigate the way that nonvanishment or vanishment is produced. The following concepts are therefore presented.

Definition 4.14. Let $\{K(\omega,t)\}_{(\omega,t)\in J\times I}$ be a function system defined in $H_1$. The set $\Lambda \subset J$ is called a frequency gap if there exists a function $f$ such that $\tilde{f}(\omega)=0$ on $\Lambda$ and $|\Lambda|\neq 0$.

On analyzing a function $f$, the transform (4.12) can be considered as a linear functional in terms of the function $f$ that acts on kernel function $K_\omega$ ($\omega \in J$) that $K(f)(\omega)=f^*(K_\omega(t))$. If this linear functional is an annihilator of the space spanned by $K_\omega$ ($\omega \in \Lambda$), then $\Lambda$ is a frequency gap. There is one of Hilbert space methods, named annihilating pairs of sets, which is employed to analysis uncertainty principle [127], [128]. Our technique partially can be understood from the view of annihilator, but it is essentially distinct. This definition is applicable to both discrete and continuous function systems. In a discrete function system, $|\Lambda|$ is a counting measure taking the size of $\Lambda$ to be the number of elements in this set if $\Lambda$ is finite, and $|\Lambda|$ is to be infinite if $\Lambda$ is an infinite set. When $\{K(\omega, t)\}$ is a CFS, $|\Lambda|$ denotes the Lebesgue measure.

Definition 4.15. Let $\{K(\omega,t)\}_{(\omega,t)\in J\times I}$ be a CFS in $H_1$, $\{J_n\}$ be a nested sequence $J \supset J_1 \supset ...J_n \supset ...$ of closed intervals and $|J_n|>0$ ($n=1, 2,...$), $\lim_{n\to\infty}|J_n|=0$. If $K(\omega, t)$ is $J_n$-complete in $H_1$ for any $n$, then $\bigcap_{n=1}^{\infty} J_n$ is called complete (function) points (abbr. as CFP or CP). The point $\omega_0$ is a simple complete point, if $\{J_n\}$ are single closed intervals so that there is a single point $\omega_0=\bigcap_{n=1}^{\infty} J_n$. Also it says $K(\omega, t)$ is complete at $\omega_0$. $S=\bigcap_{n=1}^{\infty} J_n$ is an assembly complete function point set, if it is a set of some points. $J$ is called an all-complete point set if any $\omega \in J$ is a simple complete point.

It is seen that when elements of $J_n$ ($n=1, 2,...$) are unions of intervals, their intersection can be an assembly complete function point set which is more complicated than the simple complete point. This case will be excluded most of the time. In this paper the simple complete point is mainly dealt with and is usually called a complete point simply unless otherwise specified. The complete point give an insight into the excesses of continuous function systems which cannot be developed through linear independence technique in the first part. The complete point is a global behavior of a function system rather than a local one. Some notations will be introduced to investigate it on convenience.

Notations: $\#(K)$ is the set of simple complete points of $K(\omega, t)$, $V_r(\omega_0) = \{\omega \in J: |\omega-\omega_0|<r\}$. $\chi_A$ is an indicator function of set $A$, i.e., the function whose values are 1 on $A$ and 0 elsewhere. The function $\tilde{f}(V_r(\omega_0)) = \{\tilde{f}(\omega) : \omega \in V_r(\omega_0)\}$.

A direct consequence can be obtained from above definitions and notations.

Proposition 4.16. Let $\omega_0$ be a complete point of an integral operator $K: H_1 \to H_2$ with the kernel $K(\omega, t)$. Then there doesn't exist a frequency gap $\Lambda$ containing $\omega_0$ and the corresponding function $f \in H_1$, i.e., $\tilde{f}(V_r(\omega_0)) \equiv 0$, $V_r(\omega_0) \subset \Lambda$, for any $r>0$.

This proposition is very clear. The function system is $V_r(\omega_0)$-complete, therefore there is no frequency gaps containing $\omega_0$.

Given a CFS, one wants to know whether there is a CP with this function system. Unfortunately a universal existence result of CPs cannot be established, which is easily recognized from such an example. Suppose $\{H_n(t)\}$ is the Hermite basis in $L^2(\mathbb{R})$. There is an integral kernel $K(\omega, t) = \sum_{n=0}^{\infty} \chi_{[n,n+1)}(\omega) H_n(t)$. Its integral transform is defined according to Definition 4.13. However, there is no simple complete points since functions within a small neighborhood $V_r(\omega_0) \subset [n, n+1)$ for a number $n$ and any point $\omega_0 \in \mathbb{R}^+$ are nearly all the same except $\omega_0 = n$ ($n = 1, 2, \ldots$). It is evident that the concept of complete points is in some degrees a measure of concentration of an integral kernel. It should be investigated based on the global information of them. When $\overline{\text{span}\{K(\omega,t)\}} \subset H_1$, there is information of $H_1$ not contained in $K(\omega, t)$. Thus there are no complete points of this kernel. The complete points may exist for $H_1 \subseteq \overline{\text{span}\{K(\omega,t)\}}$. In these cases, the existence of complete points will be verified given an integral operator admitting special conditions. Usually it is assumed $\overline{\text{span}\{K(\omega,t)\}} = H_1$. To present the existence results, two kinds of complete points must be firstly distinguished.

Definition 4.17. Let $K(\omega, t)$ be an integral kernel and $\omega_0$ be a complete point. If for any positive value $\varepsilon$, there is $r>0$ such that $\|K(\omega_1, t) - K(\omega_2, t)\| < \varepsilon$ for any $\omega_1, \omega_2 \in V_r(\omega_0)$. Then $\omega_0$ is called a regular CFP. Otherwise, $\omega_0$ is a singular CFP.

A complete point also can be the infinity, thus its neighborhood $V_r(\infty)$ is adopted as the form $(-\infty, -r) \cup (r, \infty)$ in which the positive value $r$ is arbitrarily large. According to Definition 4.17, the infinity as a complete point is singular.

For a finite dimensional CFS, its complete points are easy to identify from its local linear independence.

Theorem 4.18. Let $K(\omega, t)$ be an $n$-dimensional integral kernel. If its Wronskian matrix $W_n(\omega, t)$ exists and has full rank at $\omega_0$, then the point $\omega_0$ is complete.

The complete points can exist for such an integral transform when energy is conserved in the integral transform.

Theorem 4.19. Let an integral operator $K: H_1 \to H_2$ with norm preserving integral kernel $K_\omega(t) \in L^2$, $\omega \in \mathbb{R}$, $t \in I \subset \mathbb{R}$ ($I$ is a compact interval). Then the infinity is a complete point.

Proof: The new operator $K_J$ induced by the integral kernel $K(\omega, t)$ restricted on $J \times I$ ($J$ is compact) is compact, as $K(\omega, t) \in L^2(J \times I)$ and $J$, $I$ are both compact.

Suppose

$$K_J f = \tilde{f}(\omega)\chi_J(\omega) = \int_I f(t)K(\omega,t)dt, \quad \omega \in J = [-r, r], r>0.$$

While $K$ is norm preserved that $\|\tilde{f}\| = \|f\|$ and $K_J$ is a compact operator, therefore there exists a positive value $\lambda < 1$ such that

$$\| K_J f \| \leq \lambda \|f\|.$$

Here $\lambda$ can be the maximum of absolute eigenvalue of the compact operator $K_J$. But it has

$$\|\tilde{f}(\omega)\chi_{R\setminus J}(\omega)\| = \|\tilde{f}(\omega)\| - \|\tilde{f}(\omega)\chi_J(\omega)\|$$

$$= \|f\| - \| K_J f \|.$$

Hence there is

$$(1-\lambda)\|f\| \leq \|\tilde{f}(\omega)\chi_{R\setminus J}(\omega)\| \leq \|f\|. \tag{4.13}$$

where the set $\mathbb{R} \setminus J = V_r(\infty) = (-\infty, -r) \cup (r, \infty)$.

The inequality (4.13) shows that $\{K(\omega,t)\}_{(\omega,t) \in V_r(\infty) \times I}$ is a continuous frame within any neighborhood of the infinity. As a consequence, the functions of $K(\omega, t)$ at infinity is complete in $H_1$. Then the infinity is a complete point. END

When the integral operator $K$ is relaxed to be a tight continuous frame, this theorem also holds. It should be noticed that Theorem 4.19 also can be achieved when the conditions imposing on an integral kernel are altered but the induced operator $K_J$ remained compact. Besides a square-integrable kernel, there are other conditions leading to a compact integral operator, for example, a mapping $L^1 \cap L^2 \to C$ whose kernel is a continuous function of $\omega$ and $t$ [129]. Other existence results about complete points for some special integral kernels are to be presented in later part of this subsection.

*Complete Points Set*

The complete point is with the global behavior of an integral kernel. As discussed above the complete point is determined through a frequency gap and its corresponding function $f$. In fact integral kernels is directly analyzed to obtain some properties of complete points without the help of linear functional $f^*$. From the view of vector-valued functions, an integral kernel can be taken as $F(\omega) = K_\omega(t)$. The function $F$ is a vector-valued function in terms of variable $\omega$ defined on $J$ such that $F: J \to H_1$. $F(\omega)$ is said to be weakly continuous at $\omega_0$ if $\lim_{\omega \to \omega_0} f^*(F(\omega)) = f^*(F(\omega_0))$ for each $f^* \in H_1^*$, the dual space of $H_1$. Here $f^*(F(\omega)) = \langle f, F(\omega) \rangle$ with its corresponding function $f$ in $H_1$. The complete point doesn't appear occasionally and independently. Instead it is controlled by the continuity of $F(\omega)$, under which the behavior of complete points set $\#(K)$ somewhat is like the limit of a convergent sequence. Sequentially the properties of $\#(K)$ are discussed based on the continuity of integral kernels. The symbol $\#(K)$ is also denoted by $\#(F)$ sometimes interchangeably when the kernel function $K$ is associated with a vector-valued function $F$.

Theorem 4.20. Suppose the vector-valued function $F: J \to H_1$ and a point sequence $\{\omega_n\} \in \#(F)$ ($i=1, 2, \ldots$). If $\lim_{n \to \infty} \omega_n = \omega_p \in J$, then $\omega_p \in \#(F)$.

Proof: Given any $r>0$, there exists the certain value $N$ such that $\omega_n \in V_r(\omega_p)$ for all $n>N$. But there is $\delta>0$ that $V_\delta(\omega_n) \subset V_r(\omega_p)$, $n>N$. That is, there is not a function $\tilde{f}(\omega)$ can vanish on $V_\delta(\omega_n)$ and so on $V_r(\omega_p)$ for any $r>0$. While $\omega_p \in J$, then $\omega_p \in \#(F)$.    END

One can readily deduce a corollary from this theorem.

Corollary 4.21. Let $J_1 \subset J$ and $J_1 \subset \#(F)$. If its closure $\mathrm{cl}(J_1) \subset J$, then there is $\mathrm{cl}(J_1) \subset \#(F)$.

Proof: Select a sequence $\{\omega_n\} \in J_1$, there is $\lim_{n \to \infty} \omega_n \in \#(F)$ according to Theorem 4.20. Therefore the accumulation points in $J_1$ belong to $\#(F)$. But $J_1 \subset \#(F)$, it so has $\mathrm{cl}(J_1) \subset \#(F)$.    END

There is a group of complete points occurring in a continuous function system under conditions as shown above. But a complete point can also be isolated. For example, $F(\omega) = \chi_{[0,1)}(\omega)h_1(t) + \chi_{[1,2]}(\omega)h_2(t)$, $\omega \in [0, 2]$ and $H_1$ is spanned by the linearly independent functions $h_1$, $h_2$. Then it is seen that there is only the point $\omega=1$ complete as its any neighborhood contains $h_1(t)$ and $h_1(t)$, the basis of $H_1$, while the sufficiently small neighborhood of any other point has only one basis function, $h_1(t)$ or $h_1(t)$. But this complete point is singular as $F(\omega)$ is discontinuous at $\omega=1$.

To present our results, the isolated complete point should be refined. Suppose $F(\omega) = a_1(\omega)h_1(t) + a_2(\omega)h_2(t)$, $\omega \in [0, 2]$, $h_1$, $h_2$ are linearly independent and $a_1(\omega)$, $a_2(\omega)$ are continuous but vanishing on $[1, 2]$, $[0, 1]$ respectively. $F(\omega)$ is even strong continuous. Nevertheless only the point $\omega_0=1$ is complete. The characteristic of this kind of complete points is that $\tilde{f}(\omega_0)$ doesn't vanish in $V_r(\omega_0)$ but it vanishes in a set with nonzero Lebesgue measure $\Lambda \subset V_r(\omega_0)$, for any $r>0$. We call such CFPs trivial. When the continuity is commanded, there are no isolated complete points if they are not trivial.

Theorem 4.22. Suppose the vector-valued function $F(\omega) : J \rightarrow H_1$ is weak continuous. If $\omega_p \neq \infty$ is a nontrivial complete point, then the set $\#(F)$ is connected at $\omega_p$.

Proof: From the continuity of $F(\omega)$ and $\omega_p$ being a nontrivial complete point, it has $\tilde{f}(\omega) \neq 0$ in $V_r(\omega_p)$ for any $r>0$ and $f \in H_1$ except a null set. Then there exist $\delta>0$, $\omega_q \in V_r(\omega_p)$ and $V_\delta(\omega_q) \subset V_r(\omega_p)$ such that $\tilde{f}(\omega) \neq 0$ in $V_\delta(\omega_q)$ also for any $f \in H_1$ except a null set. It leads to $\omega_q$ being complete. Since the value $r$ and neighborhood $V_r(\omega_p)$ can be selected arbitrarily, that is, there always exists a CFP for any neighborhood $V_r(\omega_p)$, it means $\#(F)$ is connected at $\omega_p$.   END

As seen above, for a continuous vector-value function a nontrivial CFP is either an interior point or an endpoint and it cannot be an isolated point. These results are summarized in the following theorem.

Theorem 4.23. Let $J$ be a closed set. Then $\#(F)$ is a closed set. Specially when $F$ is continuous, the set of nontrivial complete points is the form that $\#(F)=\cup J_n$ or $\cup J_n \cup \{\infty\}$ where $J_n$ ($n=1, 2, \ldots$) are closed intervals.

Proof: For any point $\omega \in \text{cl}(\#(F))$, it is either isolated, or connected. If it is isolated, then $\omega \in \#(F)$. If it is connected, there is a point sequence $\{\omega_n\} \in \#(F)$ convergence to it. And $\{\omega_n\}$ also converges to an element in $J$, as $J$ is a closed set. Then it has $\omega = \lim_{n\to\infty} \omega_n \in J$ and $\omega \in \#(F)$ according to Theorem 4.20. It implies $\text{cl}(\#(F)) \subseteq \#(F)$.

However, #(F)⊆cl(#(F)) and so #(F)=cl(#(F)). #(F) is a closed set.

When $F$ is continuous, it is known from Theorem 4.22 there are no isolated nontrivial CFPs in #(F) except ∞. If there is a finite value $\omega \in$ #(F), then there should be an interval $J_n$ such that $\omega \in J_n \subseteq$ #(F) since it is connected also from Theorem 4.22. Note that $J_n$ should be closed, otherwise the fact #(F) a closed set is to be violated. While the point ∞ also can be complete, then #(F)= $\cup J_n$ or $\cup J_n \cup \{\infty\}$.                    END

It is clear that #(F) is also closed when it includes the point ∞. When there is not a CP within its neighborhood $V_r(\infty)$ with $r$ being sufficiently large, we call ∞ isolated.

There are some simple properties with respect to #(K) listed in the following. Denote vector-valued functions by $F$, $F_1$, $F_2$, … on $J$ and corresponding integral kernels by $K$, $K_1$, $K_2$, …. The set #($K_1$)+ #($K_2$) is exactly #($K_1$)∪#($K_2$) compatible with symbols.

Theorem 4.24. Suppose $h(t)$ be a $L^2$ function and $a$ a constant, it has

(i) #($aK$)= #($K$).

(ii) #($aK+h$)= #($K$), if $F'(\omega)$ exists and is complete at $\omega \in$ #(F).

(iii) If $\{\infty\} \in$ #($K$), then $\{\infty\} \in$ #($aK+h$).

(iv) #($K$)= #($T(K)$), if $K_\omega(t) \in H$ and $T: H \to H$ being a unitary operator.

Proof:

(i) It is clear.

(ii) Suppose $\omega_0 \in$ #($K$), $\tilde{f} =<f, K_\omega(t)>$ and $\tilde{g}=<f, aK_\omega(t)+h> = a\tilde{f} +<f, h>$. Then the function $\tilde{f}$ doesn't vanish within arbitrarily small neighborhood $V_r(\omega_0)$ for any nonzero $f \in H_1$. While, $F'(\omega)$ is complete at $\omega_0$, then it suggests $\tilde{f}$ not a constant in

$V_r(\omega_0)$. If it is not, $\tilde{f}'=<f, F'(\omega)>=0$, $\omega \in V_r(\omega_0)$. Therefore $\tilde{g}(V_r(\omega_0))$ doesn't equal 0 identically for any $f \in H_1$. It means $\omega_0 \in \#(aK+h)$ and $\#(K) \subseteq \#(aK+h)$. By the same reasoning, there is $\#(aK+h) \subseteq \#(K)$. So it has $\#(aK+h)= \#(K)$.

(iii) As $\{\infty\} \in \#(K)$, it can be proved that there is not a function $f \neq 0$ such that $\tilde{g}(\omega)=<f, aK_\omega(t)+h>\equiv$constant, $\forall \omega \in V_r(\infty)$ and any $r>0$. Otherwise $\tilde{g}, \tilde{f} \notin L^2$ and there will be a contradiction to our assumption in this paper. Hence $\{\infty\} \in \#(aK+h)$.

(iv) Suppose $\omega_0 \in \#(K)$ and it is certain that $\omega_0 \in \#(T(K))$. Otherwise, there is $f \neq 0$ and $r>0$ admitting $<f, T(K)>=0$, $\omega \in V_r(\omega_0)$. It has $<T^{-1}f, K>=<TT^{-1}f, TK>=<f, TK>=0$, which means a contradiction to the supposition of complete point $\omega_0$. Then it has $\#(K) \subseteq \#(T(K))$. Following the same way it also has $\#(T(K)) \subseteq \#(K)$. Hence there is $\#(K)= \#(T(K))$. END

One can determine complete points of a kernel function by examining a known kernel according to (iv) if the two space expanded by kernels are isomorphic. The unitary operator (isomorphism) could be freed to be a local diffeomorphism to be addressed in the later part.

*Stability of Complete Points*

The integral transform (4.12) also provides a recovery of function (signal) by its transformed function in frequency domain when the continuous frame condition is assured. Furthermore complete points indicate that this recovery is realized probably by the partial information around a complete point. A question is thus raised that under which conditions a signal can be reconstructed by partial data segments of its transform in frequency domain? Is this reconstruction numerically stable? In a very small neighborhood, the functions to be recovered may not be distinguished in numerical analysis. These problems lead us to the definition below.

Definition 4.25. Let $\omega_0$ be a complete point of an integral kernel $K(\omega, t)$ on $J \times I$ mapping from $H_1$ to $H_2$. $\omega_0$ is said to be stable if $\{K(\omega, t)| \omega \in V_r(\omega_0)\}$ constitutes a continuous frame of $H_1$ for any $r>0$. Otherwise, it is said to be unstable.

The stable complete point has a relation to a reproducing kernel Hilbert space (RKHS). If $\omega_0$ is a stable complete point, there are values $A, B >0$ for the value $r>0$ such that

$$A\|f\|^2 \leq \| \tilde{f}(\omega) \chi_{V_r(\omega_0)}(\omega)\|^2 \leq B\|f\|^2 \tag{4.14}$$

There is a synthesis continuous frame $G(t, \omega)$ reconstructing $f(t)$ from the partial information $\tilde{f}(\omega) \chi_{V_r(\omega_0)}(\omega)$ that

$$f(t) = \int_{V_r(\omega_0)} G(t,\omega) \tilde{f}(\omega) d\omega \tag{4.15}$$

and

$$\tilde{f}(\omega) = \int_I K(\omega,t) f(t) dt \tag{4.16}$$

Combine (4.15) and (4.16)

$$f(t) = \int_I R(t,\tau) f(\tau) d\tau, \tag{4.17}$$

where

$$R(t, \tau) = \int_{V_r(\omega_0)} G(t,\omega) K(\omega,\tau) d\omega. \tag{4.18}$$

The interchange of integrals in (4.17) is permissible if $\tilde{f}$ is sufficiently smooth. Therefore in this meaning, We say that $H_1$ is the reproducing kernel Hilbert space with kernel $R(t, \tau)$.

Proposition 4.26. Let $K: H_1 \to H_2$ be an integral operator with the kernel function $K(\omega,$

$t$) on $J \times I$ and $\tilde{f}$ is sufficiently smooth. If $\omega_0$ is a stable complete point, then there exists a reproducing kernel $R(t, \tau)$ of $H_1$.

The local reconstruction around stable complete points depends on the available global information of the kernel function around CFPs. Concerning types of kernels there are the following results.

Theorem 4.27. Suppose kernel $K(\omega, t)$ and its spanned space $H_1$ is finite dimensional. Its complete points are stable.

Proof: Without loss of generality, $H_1$ is assumed to be $n$ dimensional and $\omega_0$ be a complete point. One has an orthonormal basis $K_1(t), K_2(t), ..., K_n(t)$ of $H_1$ such that

$$K(\omega, t) = \sum_{i=1}^{n} \alpha_i(\omega) K_i(t), \quad \omega \in J. \tag{4.19}$$

$\alpha_i(\omega) \in L^2(J)$ ($i=1,2,...n$). $\{\alpha_i(\omega)\}$ are linear independent on $V_r(\omega_0)$ for any $r>0$. If it is not, then there is $f \in H_1$ such that $<f, K(\omega, t)> = \sum_{i=1}^{n} c_i \alpha_i(\omega) = 0$, $\omega \in V_r(\omega_0)$, $c_i = <f, K_i(t)>$ ($i=1,2,...n$). But $K(\omega, t)$ posses at least a complete point. So it is impossible that $\{\alpha_i(\omega)\}$ are linearly dependent on $V_r(\omega_0)$.

Hence it has

$$\tilde{f}(\omega) \chi_{V_r(\omega_0)}(\omega) = <f(t), K(\omega, t) \chi_{V_r(\omega_0)}(\omega)>$$

$$= \sum_{i=1}^{n} \alpha_i(\omega) <f(t), K_i(t)> \chi_{V_r(\omega_0)}(\omega)$$

$$= \sum_{i=1}^{n} c_i \beta_i(\omega). \tag{4.20}$$

where $\beta_i(\omega) = \alpha_i(\omega) \chi_{V_r(\omega_0)}(\omega)$ ($i=1,2,...n$).

The energy of function $\tilde{f}$ around $\omega_0$ is

$$\|\tilde{f}(\omega)\chi_{V_r(\omega_0)}(\omega)\|=\|\sum_{i=1}^{n}c_i\beta_i(\omega)\|=(cAc^T)^{1/2}, \qquad (4.21)$$

$$c=(c_1, c_2, ..., c_n) \text{ and } A=\begin{bmatrix} <\beta_1(\omega),\beta_1(\omega)> & \cdot & \cdot & \cdot & <\beta_1(\omega),\beta_n(\omega)> \\ & \cdot & & & \cdot \\ & \cdot & & & \cdot \\ & \cdot & & & \cdot \\ <\beta_n(\omega),\beta_1(\omega)> & \cdot & \cdot & \cdot & <\beta_n(\omega),\beta_n(\omega)> \end{bmatrix}.$$

It is known that $A$ is a Gramian matrix; its determinant is not singular as $\{\beta_i(\omega)\}$ are linear independent. Therefore there are two values $\lambda_M \geq \lambda_m > 0$ such that

$$\lambda_m^{1/2}\|c\| \leq (cAc^T)^{1/2} \leq \lambda_M^{1/2}\|c\|. \qquad (4.22)$$

While $K_1(t), K_2(t),..., K_n(t)$ are normalized orthogonal basis, so

$$\|f\|^2 = \sum_{i=1}^{n}|<f,K_i>|^2 = \|c\|^2 \qquad (4.23)$$

Replace $c$ in (4.22) with (4.23),

$$\lambda_m\|f\|^2 \leq \|\tilde{f}(\omega)\chi_{V_r(\omega_0)}(\omega)\|^2 \leq \lambda_M\|f\|^2. \qquad (4.24)$$

It implies $K(\omega, t)$ on $\chi_{V_r(\omega_0)} \times I$ being a continuous frame. Then $\omega_0$ is a stable complete point.                 END

Although the local reconstruction around a CP is stable for the finite dimensional kernel, the performance of numerical reconstruction can become degenerated if the neighborhood $V_r(\omega_0)$ is very small. For an infinite dimensional integral kernel, its stability of a complete point has rich behaviors than its finite counterpart. An intuitive idea of stable complete points for an infinite dimensional system is that it is required of enough functions belonging to this kernel to locate in an infinite interval. There is indeed the following theorem supporting this viewpoint.

Theorem 4.28. A regular complete point of an infinite dimensional integral kernel is unstable.

Proof: Suppose the integral kernel $K(\omega, t)$ on $J \times I$ and the complete point is $\omega_0$. The integral kernel restricted in a neighborhood $V_r(\omega_0)$ is

$$\int_{V_r(\omega_0)} \int_I |K(\omega,t)|^2 \, d\omega dt \leq \int_{V_r(\omega_0)} \max_{\omega \in V_r(\omega_0)} \|K(\omega,t)\|^2 d\omega$$
$$= V_r(\omega_0) \max_{\omega \in V_r(\omega_0)} \|K(\omega,t)\|^2 < \infty \qquad (4.25)$$

In the above, $|V_r(\omega_0)|$ is finite because $\omega_0$ is regular which cannot be $\infty$. Its associated integral operator is then compact from (4.25). There exists $\{\lambda_n\}$ which tends to zero such that

$$\int_I K(\omega,t)\chi_{V_r(\omega_0)}(\omega) f_n(t) dt = \lambda_n f_n \qquad (4.26)$$

Hence there is $\|\tilde{f}_n(\omega) \chi_{V_r(\omega_0)}(\omega)\|/\|f_n\|$ tends to zero. It means $\{K(\omega, t) | (\omega, t) \in V_r(\omega_0) \times I\}$ doesn't constitute a continuous frame in its spanned space. Then the regular complete point is unstable. END

It has been known the infinity is a complete point of a norm preserving integral operator, and the integral kernel within the neighborhood of this CP is a continuous frame from the proof of Theorem 4.19. The stability of this complete point is thus readily obtained.

Theorem 4.29. Let $K: H_1 \rightarrow H_2$ be an norm preserving integral operator with the kernel function $K(\omega, t)$ on $\mathbb{R} \times I$. $I$ is a finite interval and $K_\omega(t) \in L^2(I)$ for any $\omega \in \mathbb{R}$. Then the infinity is a stable singular complete point.

This theorem has an application in signal area to time-and bandlimited functions which can be understood from this view of stability of the infinity complete point.

Example 4.30. In signal analysis, spectrum estimation of a signal is of importance and the Fourier transform is often applied for this pursuit. However, a signal in the real world is always timelimited. Therefore the Fourier transform deals with a truncated signal and is associated with an integral kernel $\{e^{-i\omega t}\}_{(\omega,t)\in\mathbb{R}\times I}$ ($I$ is a finite support). The well known result about it, also viewed as one formulation of the qualitative uncertainty principle, is that the transformed function doesn't vanish at the infinity. It is actual a deduction from the fact that $\{e^{-i\omega t}\}$ around $\omega=\infty$ composing a continuous frame. The prolate spheroidal wave function $f(t)$ timelimited on $I$ has the maximal energy concentration over a finite frequency support $J$. Denote this maximal value by $\lambda_0$ that

$$\lambda_0 = \frac{\int_J |\tilde{f}(\omega)|^2 d\omega}{\int_{-\infty}^{\infty} |\tilde{f}(\omega)|^2 d\omega} \quad (4.27)$$

Then

$$(1-\lambda_0)\|f\| \leq \|\tilde{f}(\omega)\chi_{\mathbb{R}\setminus J}(\omega)\| \leq \|f\| \quad (4.28)$$

The above provides us a reconstruction scheme of timelimited functions without necessary the whole information over the frequency range but only preserving the frequency information in the infinity. But we see that the lower bound $(1-\lambda_0)$ of (4.28) tends to zero very quickly when $J$ is getting large. The reconstruction will become worse in numerical computation. On the other hand, Slepian argued [90] that in the communication theory a real world signal is actually a time-and bandlimited function, where the timelimited function is approximated by the obtained information only in finite frequency band. It means an exact reconstruction impossible.

The stable complete point theory may be attractive and fascinating for the description of signals in a very limited data space. "In small proportions we just beauties see; and in short measures life may perfect be"[*]. But the theorem about stability of regular complete point gives a negative answer and the infinity CP also indicates there should be a measure covering an infinite interval. We will see in the example followed that singular complete points located at the finite can serve us this ideal purpose. Unfortunately it is not physically realizable.

Example 4.31. Considering the kernel $\{e^{it/\omega}\}_{(\omega,t)\in\mathbb{R}\times I}$, it is clear that $\omega=0$ is a complete point just corresponding to $\omega=\infty$ in $\{e^{i\omega t}\}_{(\omega,t)\in\mathbb{R}\times I}$. With the information of a frequency signal in a neighborhood about $\omega=0$, it is enough to reconstruct the original signal in time domain. But the kernel is singular at this complete point and in practice the exact frequency information around 0 is also unavailable because it has the data segment with infinite length. It makes the reconstruction a highly ill-posed problem and even impossible really.

*B2. Geometry Theory of Complete Points*

The complete point is characterized by behaviors of the neighborhood of this point in the previous. But this technique is cumbersome and doesn't supply a clear insight into the complete point. Maybe one wants some quantities directly related to a CP itself, while not to address its vicinity every time. In this subsection, integral kernels are allowed to deal with differentiability and thus are considered as differentiable manifolds in Hilbert space. Generally the study of complete points is reduced to studies of full rank of a map between two manifolds. Especially when one manifold is selected as the nature Euclidean space, the identification of complete points of integral kernels becomes retrieval of regular points on the manifold. In this geometric setting, complete points theory of finite, one and multi-dimensional integral kernels can be clearly understood.

---
[*] A quote from the poem "The Noble Nature" by Ben Jonson

If order of differentiability is enough imposed on the map, then complete points can be studied with Taylor's formula as introduced below.

Theorem 4.32 [130]. If a mapping $F: M \rightarrow N$ from a neighborhood $V=V(\omega)$ of a point $\omega$ in a normed space $M$ into a normed space $N$ has derivatives up to order $n$-1 inclusive in $M$ and has an $n$-th order derivative $F^{(n)}(\omega)$ at the point $\omega$, then

$$F(\omega+h) = F(\omega) + F'(\omega)h + \ldots + F^{(n)}(\omega)h^n / n! + o(|h|^n)$$

as $h \rightarrow 0$.

A vector-valued function $F: J \rightarrow H_1$ is a special case of mappings from $M$ to $N$. As a consequence, complete points are recognized by their derivatives.

Theorem 4.33. If the vector-valued function $F(\omega)$ is $n$-dimensional, $F^{(i)}(\omega)$ ($i=0, \ldots, n$-1) exists and they are linear independent in a neighborhood about $\omega_0$, then $\omega_0$ is a CP.

Proof: When $n=1$, the theorem is trivial. Suppose the theorem is true when $n=k$, $F^{(i)}(\omega)$ ($i=0, \ldots, k$-1) exists and are linear independent in a neighborhood about $\omega_0$. Then when $n=k+1$ the derivatives of $F^{(1)}(\omega)$ exists at $\omega_0$ up to $k$-1 order and $\{F^{(1)}(\omega), F^{(2)}(\omega), \ldots, F^{(k)}(\omega)\}$ are linear independent in a neighborhood about $\omega_0$. $F^{(1)}(\omega)$ is complete at $\omega_0$ according to our supposition in the case $n=k$. It will be shown $F(\omega)$ is complete at $\omega_0$. Otherwise there is a differentiable function $\tilde{f}(\omega)$ ($\omega \in V(\omega_0)$) vanishing in a neighborhood $V(\omega_0)$. It has $\tilde{f}'(\omega)=<f, F^{(1)}(\omega)>=0$ for $\omega \in V(\omega_0)$. But $\omega_0$ is a CP of $F^{(1)}(\omega)$. Therefore $\omega_0$ is a CP of $F(\omega)$.     END

This theorem is an abstraction of Theorem 4.18. It is necessary that $F^{(i)}(\omega)$ should not vanish in a neighborhood of CP. For some classes of kernels, the full rank of derivatives at a point may also provide the result that this point being complete. This study can be established on manifolds which set a framework for investigation of complete points. In the context of manifolds the CFS can be extended to multidimensional index.

Definition 4.34. Let $F$: $J \to H_1$ be a vector-valued function, $J \subseteq \mathbb{R}^n$, $H_1 \subseteq L^2(\mathbb{R})$. There is $\boldsymbol{\omega} \in J$ denoted by $\boldsymbol{\omega}=(\omega^1,\ldots, \omega^n)$ in the conventional terminology in geometry. $F(\boldsymbol{\omega})$ is called a multi-index function and its corresponding integral kernel is $K(\omega^1,\ldots, \omega^n, t)= F(\boldsymbol{\omega})$.

In computational harmonic analysis, there are some integral transforms, for example, Wigner distribution, wavelet and recent rigdelet and curvelet transforms, mapping $L^2(\mathbb{R})$ to $L^2(\mathbb{R}^n)$. The multi-index function has a geometric structure which provides a framework analyzing these transforms with the complete point theory. To serve our purpose, it is required of some necessary basic notions in differential geometry [131].

Defintion 4.35. Let $M$, $N$ be the normed linear spaces or manifolds and the mapping $F$: $M \to N$ is a bijective, differentiable operator, and so its inverse $F^{-1}$: $N \to M$, then $F$ is called a diffeomorphism between $M$ and $N$. A diffeomorphism defined on some open neighborhood $V$ of a given point $x$ is called a local diffeomorphism at $x$. Local diffeomorphisms at *x* can be regarded simply as invertible changes of local coordinates near *x*.

The symbol *F* will be employed denoting mapping between manifolds (spaces) or abstract functions without confusion. There is an important result regarding diffeomorphisms and derivative of the map.

Theorem 4.36. If $F: M \to N$ is a local diffeomorphism at *x* in *M* then $dF_x : T_xM \to T_{F(x)}N$ is a linear isomorphism. Conversely, if $dF_x$ is an isomorphism then there is an open neighborhood *V* of *x* such that *F* maps *V* diffeomorphically onto its image.

When the mapping is not a diffeomorphism, there are two possible kinds of points in the manifold.

Definition 4.37. Let *M* and *N* be differentiable manifolds and $F: M \to N$ be a differentiable map between them. If its differential $dF_x : T_xM \to T_{F(x)}N$ is a surjective linear map at a point $x \in M$, i.e. the rank of the map $dF_x$ is equal to dim *N*. In this case *x* is called a regular point of the map *F*, otherwise, *x* is a singular point. A point $y \in N$ is a regular value of *F* if all points *x* in the pre-image $F^{-1}(y)$ are regular points.

It is enough for us to convert integral kernels to manifolds. Consider an integral kernel $K(\omega^1,\ldots, \omega^m, t)$, its definitive Hilbert space $H_1$. Select an orthonormal basis $\{\varphi_i(t)\}$, $K_\omega(t)$ has an expansion that $K_\omega(t) = \sum_i \psi_i(\omega^1,\ldots,\omega^m)\varphi_i(t)$. To make this expansion meaningful, $\sum_i \psi_i(\omega^1,\ldots,\omega^m)\varphi_i(t)$ with respect to each $\omega$ converges to $K_\omega(t)$ in $L^2$ norm sense when it is infinite dimensional. Further, a one-to-one correspondence can be established between $\{\varphi_i(t)\}$ and $\{e_i\}$, the natural basis of Euclidean space. The space $H_1$ is isomorphic to Euclidean space $\mathbb{R}^n$ (*n* is alternatively infinite) for the certain value $n$. Therefore the kernel $K(\omega^1,\ldots, \omega^m, t)$ ($m\leq n$) is actually an $m$-dimensional manifold parameterized by $(\omega^1,\ldots, \omega^m)\in \mathbb{R}^m$ that $M=(\psi_1(\omega^1,\ldots, \omega^m), \ldots, \psi_n(\omega^1,\ldots, \omega^m))$, specially $M=(\psi_1(\omega), \ldots, \psi_n(\omega))$ for usual two dimensional integral kernel is a space curve. The notion of dimensionality here refers to different contents. One refers to the manifold $M$ with $m$ dimension; the other refers to the Hilbert space $H_1$ with $n$ dimension. We will distinguish them in the specific statement.

Firstly we will set up the relation of complete points between integral kernels and manifolds.

Definition 4.38. Let $f$ be a nonzero scale function associated with a vector also denoted by $f \in \mathbb{R}^n$ mapping from manifold $M = (\psi_1(\omega), \ldots, \psi_n(\omega))$ to the real axis $\mathbb{R}$ such that $f(x)=<f, x>$ at each point $x\in M$. If for any small neighborhood $V(x)$, $f(V(x))=0$ only induces $f=0$, then $x$ is a complete point.

When $x$ is a complete point, there are points $\{x_1, \ldots, x_n\}$ in any neighborhood $V(x)$ such that the group of vectors $\{xx_1, \ldots, xx_n\}$ or $\{x_1, \ldots, x_n\}$ spans $\mathbb{R}^n$ and vice versa. We will see the definition of complete points here is equivalent to that in integral kernels clearly from the parameterization process of manifolds from integral kernels.

Proposition 4.39. Given an integral kernel $K(\omega, t)$ and its corresponding parameterized manifold $M$, $\omega$ is a complete point associated with the point $x$ in the manifold. Then $x$ is a complete point of $M$, and vice versa.

The differentiable mapping between two manifolds, also two integral kernels will be presented extending notion of unitary operators in (iv) of Theorem 4.24.

Theorem 4.40. Given two differentiable manifolds $M$ and $N$, dim $M$=dim $N$, their corresponding Hilbert spaces $H_M = H_N$ and $F : M \to N$ is a differentiable map between them. If the point $x_0$ of $M$ is regular of the map $F$ and $x_0 \in \#(M)$, then $F(x_0) \in \#(N)$.

Proof: For $F$ being a differentiable map, $F(V(x_0))$ is also a neighborhood of $F(x_0)$. If $F(x_0) \notin \#(N)$, there is a nonzero scale function $f$ such that $f(F(x))=0$ for any $x \in V(x_0)$ and a neighborhood $V(x_0)$. Differentiating two sides of the above equality, then it has $df(F(x))/dx=0$ at $x_0$. But $df(F(x))/dx= (f_{F(x)}) \cdot dF_x = f^T dF_x$. Here $f^T = (f_1, \ldots, f_n)$, $n=$ dim $N$. Because $x_0$ is regular, $dF_x$ is full rank at $x_0$, then $df(F(x))/dx=(f_{F(x)}) \cdot dF_x \neq 0$ at $x= x_0$. It is contradictory to our supposition. Therefore there is $F(x_0) \in \#(N)$.         END

When the differentiable map $F$ is one-to-one, the complete points $x_0$ of $M$ and $F(x_0)$ of $N$ actually correspond to the same parameter $\omega_0$ in the integral kernel. In this regard, we write $x_0 \sim F(x_0)$. As a consequence, there are two corollaries which we just state without proofs.

Corollary 4.41. Given a local diffeomorphism $F$ from $M$ to $N$, it has $\#(M) \sim \#(N)$.

Corollary 4.42. Let $M$ and $N$ be two differentiable manifolds. There is an induced tangent map $dF_x : T_x M \to T_{F(x)} N$. If it is full rank at $x_0$, and $x_0 \in \#(M)$, then $F(x_0) \in \#(N)$.

The two corollaries present concrete results of comparison of two complete point sets of different kernels. When we investigate one manifold $N$ individually, the Euclidean space will play the role like the pre-image $M$.

After the general results about complete points in manifolds, we will focus on two dimensional integral kernels which are mostly addressed. The kernels are related to space curves geometrically. First two definitions are necessary.

Definition 4.43. A $C^r$-curve $\gamma$: [a, b] $\to \mathbb{R}^n$ is called regular of order $m$ if $\{\gamma^{(1)}(\omega), \gamma^{(2)}(\omega), \ldots, \gamma^{(m)}(\omega)\}$, $m \leq n$ are linearly independent in $\mathbb{R}^n$. If $m$ is infinite, we call it regular of infinite order.

The Frenet frame of a smooth curve denoted by $e_1(\omega), \ldots, e_n(\omega)$, is constructible from derivatives of $\gamma(\omega)$ using Gram-Schmidt orthogonalization algorithm. A generalized curvature can be defined from the frame.

Definition 4.44. The functions $\lambda_i(\omega)$ are called generalized curvatures, defined as

$$\lambda_i(\omega) = <e'_i(\omega), e_{i+1}(\omega)> / \|\gamma'(\omega)\| \tag{4.29}$$

where $\|\gamma'(\omega)\|$ is the magnitude of the tangent vector.

The complete point is in some sense determined by the regular order of a curve and its generalized curvatures. The two theorems will show it.

Theorem 4.45. Given a smooth $n$-dimensional curve $\gamma$: $I \to \mathbb{R}^n$, if $x_0$ associated with the parameter $\omega_0$ is a complete point, then $\gamma$ is regular of order $n$ at $x_0$ (or in the vicinity of $x_0$), and vice versa.

Proof: Because of smoothness of $\gamma$, the regularity at $x_0$ is meant the regularity in its vicinity. If the point $x_0$ is complete but $\gamma$ is regular of order less than $n$, then $\{\gamma^{(1)}(\omega), \gamma^{(2)}(\omega), \ldots, \gamma^{(n)}(\omega)\}$ are linearly dependent in vicinity of $x_0$. And $\gamma^{(i)}(\omega)$ ($i>n$) are expressible by $\{\gamma^{(1)}(\omega), \gamma^{(2)}(\omega), \ldots, \gamma^{(n)}(\omega)\}$. The curve $\gamma(\omega)$ can be expanded in the neighborhood of $\omega_0$. It has

$$\gamma(\omega) - \gamma(\omega_0) = \gamma^{(1)}(\omega_0)(\omega-\omega_0) + \frac{1}{2!}\gamma^{(2)}(\omega_0)(\omega-\omega_0)^2 + \ldots + \frac{1}{n!}\gamma^{(n)}(\omega_0)(\omega-\omega_0)^n + \ldots \tag{4.30}$$

$$= \gamma^{(1)}(\omega_0)\alpha_1(\omega) + \gamma^{(2)}(\omega_0)\alpha_2(\omega) + \ldots + \gamma^{(n)}(\omega_0)\alpha_n(\omega) \tag{4.31}$$

Here $\{\alpha_i(\omega)\}$ are power series in terms of $\omega$ and they are linearly independent known from their constitutions. Set vectors $x_i x_0 = \gamma(\omega_i) - \gamma(\omega_0)$ ($i=1, \ldots, n$) for arbitrary selection of points $\{x_1, x_2, \ldots, x_n\}$. From (4.31) and the local linear dependence of $\{\gamma^{(1)}(\omega), \gamma^{(2)}(\omega), \ldots, \gamma^{(n)}(\omega)\}$ in vicinity of $x_0$, these vectors $\{x_i x\}$ are also linearly dependent. That is, the curve $\gamma$ is not complete at $x_0$, which violates our supposition. Therefore it is proved that the complete point indicates "full order" regular curve $\gamma$ at this point. Conversely, if $\gamma$ is regular of order $n$ at $x_0$, then $\{\gamma^{(1)}(\omega), \gamma^{(2)}(\omega), \ldots, \gamma^{(n)}(\omega)\}$ are also linearly independent in $V(x_0)$. And there is a set $\{\omega_i\}$ ($i=1, \ldots, n$) and its corresponding vectors $\{x_i x\}$ linear independence according to (4.31). It implies $\omega_0$ is complete. END

The regular order of a curve links the generalized curvatures, which is more clearly describing a complete point than the regular order.

Theorem 4.46. Given a $C^{n+1}$ $n$-dimensional space curve $\gamma$, the point $x_0$ is a complete point if and only if $\lambda_i(\omega_0) \neq 0$ ($i=1, \ldots, n-1$), $\omega_0$ is the parameter value associated with $x_0$.

Proof: If $\lambda_i(\omega_0) \neq 0$ for $i=1, \ldots, n-1$, then there is Frenet frame $\{e_1(\omega), \ldots, e_n(\omega)\}$ of curve $\gamma(\omega)$ at $\omega_0$. That is, $\gamma(\omega)$ is regular of order $n$ with this local frame in a neighborhood $V(\omega_0)$. So $x_0$ is a complete point according to Theorem 4.45.

The necessary part of this proof: Suppose $x_0$ is complete, then $\gamma(\omega)$ is regular with order $n$, and $\{\gamma^{(1)}(\omega), \gamma^{(2)}(\omega), \ldots, \gamma^{(n)}(\omega)\}$ are linearly independent in any neighborhood $V(\omega_0)$ from Theorem 4.45. Hence $\{\gamma^{(2)}(\omega), \gamma^{(3)}(\omega), \ldots, \gamma^{(n+1)}(\omega)\}$ are also locally linear independent. Assumed Frenet frame $(e_1(\omega), \ldots, e_n(\omega))^T = T(\gamma^{(1)}(\omega), \gamma^{(2)}(\omega), \ldots, \gamma^{(n)}(\omega))^T$, also there is $(e_1^{(1)}(\omega), \ldots, e_n^{(1)}(\omega))^T = T(\gamma^{(2)}(\omega), \gamma^{(3)}(\omega), \ldots, \gamma^{(n+1)}(\omega))^T$. $T$ is a lower triangular matrix from Gram-Schmidt orthogonalization process.

While, from the representations of generalized curvatures (4.29) by setting $\|\gamma'(\omega)\|=1$ without loss of generality, it has the Frenet–Serret formulas in $n$ dimensions such that

$$\begin{bmatrix} e_1^{(1)}(\omega) \\ e_2^{(1)}(\omega) \\ . \\ . \\ . \\ e_n^{(1)}(\omega) \end{bmatrix} = \begin{bmatrix} 0 & -\lambda_1(\omega) & 0 & . & . & . & 0 \\ \lambda_1(\omega) & 0 & -\lambda_2(\omega) & . & . & . & 0 \\ 0 & \lambda_2(\omega) & 0 & . & . & . & . \\ . & . & . & . & . & . & . \\ . & . & . & . & . & . & . \\ . & . & . & . & . & . & -\lambda_{n-1}(\omega) \\ 0 & 0 & . & . & . & \lambda_{n-1}(\omega) & 0 \end{bmatrix} \begin{bmatrix} e_1(\omega) \\ e_2(\omega) \\ . \\ . \\ . \\ e_n(\omega) \end{bmatrix} \quad (4.32)$$

Replacing $\{e_1(\omega), \ldots, e_n(\omega)\}$ in (4.32) with derivatives of $\gamma$, it becomes

$$T \begin{bmatrix} \gamma^{(2)}(\omega) \\ \gamma^{(3)}(\omega) \\ . \\ . \\ . \\ \gamma^{(n+1)}(\omega) \end{bmatrix} = \begin{bmatrix} 0 & -\lambda_1(\omega) & 0 & . & . & . & 0 \\ \lambda_1(\omega) & 0 & -\lambda_2(\omega) & . & . & . & 0 \\ 0 & \lambda_2(\omega) & 0 & . & . & . & . \\ . & . & . & . & . & . & . \\ . & . & . & . & . & . & . \\ . & . & . & . & . & . & -\lambda_{n-1}(\omega) \\ 0 & 0 & . & . & . & \lambda_{n-1}(\omega) & 0 \end{bmatrix} T \begin{bmatrix} \gamma^{(1)}(\omega) \\ \gamma^{(2)}(\omega) \\ . \\ . \\ . \\ \gamma^{(n)}(\omega) \end{bmatrix}$$

$= AT(\gamma^{(1)}(\omega), \gamma^{(2)}(\omega), \ldots, \gamma^{(n)}(\omega))^T.$ \hfill (4.33)

Comparing two sides of (4.33), the matrix $A$ should be invertible leading to $\lambda_1(\omega)\neq 0$ and $\lambda_{n-1}(\omega)\neq 0$. The further analysis also shows $\lambda_i(\omega)\neq 0$ for $i=2, \ldots, n-2$. Otherwise for any value $i=2, \ldots, n-2$, supposing $\lambda_i(\omega)=0$, the $i$-th element of left side in (4.33), noted as $l_i$, is $\Sigma(\gamma^{(2)}(\omega), \ldots, \gamma^{(i+1)}(\omega))$ (linear combination of $\{\gamma^{(2)}(\omega), \ldots, \gamma^{(i+1)}(\omega)\}$) because $T$ is a lower triangular matrix. And the corresponding element $r_i$, of right side in (4.33) is $\lambda_{i-1}(\omega) \Sigma(\gamma^{(1)}(\omega), \ldots, \gamma^{(i-1)}(\omega))$ (linear combination of $\{\gamma^{(1)}(\omega), \ldots, \gamma^{(i-1)}(\omega)\}$). From equality (4.33) it has $l_i = r_i$ suggesting linear dependence of $\{\gamma^{(1)}(\omega), \ldots, \gamma^{(i+1)}(\omega)\}$. But it's impossible from our assumption. Then it has $\lambda_i(\omega) \neq 0$ for all $i=1, \ldots, n-1$.   END

With discrimination rules of the linear independence of functions in first part of this section, a complete point is recognized by the determinant of derivatives of a curve.

Corollary 4.47. Given a $C^{n+1}$ space curve, $\gamma = (\gamma_1(\omega), \ldots, \gamma_n(\omega))$, the sufficient and necessary condition of $\omega_0$ being a complete point in $\mathbb{R}^n$ is that Wronskian determinant $W(\omega) = |\gamma^{(1)}(\omega)\ \gamma^{(2)}(\omega) \ldots \gamma^{(n)}(\omega)|$ doesn't vanish in vicinity of $\omega_0$.

The generalized curvatures are the extension of notions such as curvature, torsion of curves in three-dimensional space. Intuitively, curvature is an amount by which the "straightness" of a curve is measured; the torsion of a curve measures how sharply it is twisting, or in some degrees difference from a plane curve. The consideration of generalized curvatures with complete points is exactly development from such an intuitive idea about three-dimensional curves. The vanishments of generalized curvatures ensure its neighborhood of a point doesn't be contained in a subspace of $\mathbb{R}^n$, therefore ensure it a complete point. However, this geometrical theory is not so evidently transferred to an infinite dimensional space. In infinite case a point that is to be complete should be regular of infinite order at least. Other extensive results are too complicated to be addressed here.

*B3. Some Special Integral Kernels*

In this part some kernels, namely, Hilbert-Schmidt kernels, reproducing kernels and translation invariant kernels, which constitute a large class of usual integral transforms, are investigated in more detail to show their special properties.

*Hilbert-Schmidt Kernels*

A Hilbert-Schmidt (simplified as HS) kernel is a function $K(\omega,t)$ of $L^2(J \times I)$ that

$$\int_J \int_I |K(\omega,t)|^2 \, d\omega dt < \infty.$$

The associated Hilbert-Schmidt integral operator is the operator $K: H_1 \subseteq L^2(I) \to H_2 \subseteq L^2(J)$, an extension of a finite rank operator. This HS operator is continuous and compact. It has a concise spectral representation admitting the so-called singular value decomposition (SVD), which is also applicable to other compact operators, that $K(\omega,t) = \sum_{i=1}^{\infty} \lambda_i \psi_i(\omega) \varphi_i(t)$. The tuples $(\lambda_i; \psi_i, \varphi_i)$ is the singular system such that $\{\psi_i\}$, $\{\varphi_i\}$ are biorthogonal systems and $\sum_{i=1}^{\infty} |\lambda_i|^2 < \infty$.

From the SVD of HS operator, it has $\tilde{f} = Kf = \sum_{i=1}^{\infty} \lambda_i <f, \varphi_i> \psi_i$. Set $a_i = \lambda_i <f, \varphi_i>$, $\tilde{f} = \sum_{i=1}^{\infty} a_i \psi_i$. Note that $\{<f, \varphi_i>\}$, $\{\lambda_i\} \in l^2$, and it yields $\{a_i\} \in l^2$. The image $H_2$ of HS operator $K$ is thus spanned by $\{\psi_i\}$. And the frequency gap of $\tilde{f}$ is actually determined by the local linear dependence of $\{\psi_i\}$.

Proposition 4.48. For an HS kernel, $\omega_0$ is a complete point if and only if $\{\psi_i(\omega)\}_{i=1}^{\infty}$ are of local linear independence at $\omega_0$.

Proof: If $\{\psi_i(\omega)\}_{i=1}^{\infty}$ are of local linear independence at $\omega_0$, i.e., linearly independent in a small neighborhood $V_r(\omega_0)$ for $r>0$, then $\tilde{f} = \sum_{i=1}^{\infty} \lambda_i <f, \varphi_i> \psi_i$ doesn't vanish in $V_r(\omega_0)$ for any $f \in H_1$. It implies $\omega_0$ a complete point.

Conversely if $\omega_0$ is complete, $\tilde{f} = \sum_{i=1}^{\infty} a_i \psi_i$ doesn't vanish over $V_r(\omega_0)$ for any

$\{a_i\} \in l^2$ and $r>0$. It means $\{\psi_i(\omega)\}_{i=1}^{\infty}$ are of local linear independence at $\omega_0$.  END

It is known that $I$ and $J$ are both finite supports, the HS kernel will not have a stable CP according to Theorem 4.28. However, HS kernels even don't have a CP under some strict conditions. When $\{\psi_i\}$ spans the space $L^2(J)$, any function in $L^2(J)$ can be expanded by $\{\psi_i\}$. In this case there is no CPs existed. Following the reverse way, we consider such questions: whether the superposition of a sequence $\{\psi_i\}$ vanishing on a neighborhood of point $\omega \in J$ indicates this sequence completeness in $L^2(J)$? $\{\psi_i\}$ that is incomplete in $L^2(J)$ has at least a complete point in $J$? A sequence $\{\psi_i\}$ can be linearly independent on any interval? These questions are answered in the following two examples.

Example 4.49. Consider a trigonometric sequence $\{\cos n\omega\}_{n=0}^{\infty}$ in $L^2[-\pi, \pi]$. An even function $f \in L^2[-\pi, \pi]$ is a linear combination of $\{\cos n\omega\}$. It is certain that $f$ can be selected vanishing on any pre-assigned support in $[0, \pi]$ or $[-\pi, 0]$. But $\{\cos n\omega\}$ is incomplete in $L^2[-\pi, \pi]$ and also there are no complete points in $[-\pi, \pi]$ if the HS kernel is with $\psi_i(\omega)=\cos(i\omega)$. Thus it is disproved the first two problems.

Example 4.50. The power sequence $\{1, \omega, \omega^2, ...\}$ in $C[0,1]$ is a basis for analytic functions that $f(\omega)=f(\omega_0)+f^{(1)}(\omega_0)(\omega-\omega_0)+ ...$ . To make $f(\omega)$ be zero at points around $\omega_0$, the sequence $\{f(\omega_0), f^{(1)}(\omega_0), ...\}$, which are coefficients of the power series, should be null. Then it is proved that an infinite function sequence can be linearly independent on any finite interval.

As seen above, the linear independence doesn't lead to completeness of a function sequence directly in its defined space. A trivial result is that if $\{\psi_i\}$ are linear independent on $V(\omega)$, then they are of linear independence on $J \supseteq V(\omega)$. We call the former local independence, and the latter global independence. That is, the local independence implies global independence while the inverse doesn't work. In order to address the relation between local independence and CPs, a new concept is introduced.

Definition 4.51. Given a linear independent DFS $\{\psi_i(\omega)\}$ on $J \subseteq \mathbb{R}$, it is defined $I(\omega)=\inf\{r: \{\psi_i(\omega)\}$ are linearly independent in the neighborhood $V_r(\omega)$ of $\omega\}$. $I(\omega)$ is called the independent radius of a point.

Obviously if $\{\psi_i(\omega)\}$ is of linear dependence on $\mathbb{R}$, this definition isn't well applied to them. In this case $I(\omega)$ cannot get any value. Based on the above definition, there are sequentially some properties about independent radius.

Proposition 4.52. $I(\omega)$ is defined on $J \subseteq \mathbb{R}$. It has

(i) $I(\omega) \geq 0$.

(ii) $|I(\omega_i) - I(\omega_j)| \leq |\omega_i - \omega_j|$.

(iii) $I(\omega)$ is continuous.

(iv) $I(\omega_0)=0$ if and only if $\omega_0$ is complete.

(v) The complete set is closed.

Proof: (i) It's trivial.

(ii) Set $r_1 = I(\omega_i)+|\varepsilon|$ where $\varepsilon$ is an infinitesimal quantity and $\{\psi_i(\omega)\}$ are linearly independent on $V_{r_1}(\omega_i)$. For another point $\omega_j \neq \omega_i$, $r_2$ is properly chosen that $V_{r_1}(\omega_i)$ is contained within $V_{r_2}(\omega_j)$ exactly, i.e. $V_{r_1}(\omega_i) \subset V_{r_2}(\omega_j)$. It has $r_2=|\omega_i - \omega_j|+r_1$, which leads to a new inequality that $r_2 - r_1 \leq |\omega_i - \omega_j|$. But $I(\omega_j) \leq r_2$ because $\{\psi_i(\omega)\}$ are linearly independent on $V_{r_1}(\omega_i)$ thus on $V_{r_2}(\omega_j)$. While, $r_1= I(\omega_i)+|\varepsilon|$, there is $|I(\omega_i)- I(\omega_j)|\leq r_2-r_1+|\varepsilon|=|\omega_i- \omega_j|+|\varepsilon|$. Let $\varepsilon$ tends to zero, it has $|I(\omega_i)- I(\omega_j)|\leq|\omega_i- \omega_j|$.

(iii) For any given $\varepsilon>0$, one select $\delta$ such that $|\omega- \omega_0|<\delta<\varepsilon$. From (ii), it has $|I(\omega)-I(\omega_0)|\leq|\omega-\omega_0|<\delta<\varepsilon$. $I(\omega)$ is continuous and even Lipschitz continuous.

(iv) It's trivial.

(v) Set $K$ an corresponding HS kernel with respect to $\{\psi_i(\omega)\}$. When the complete point is isolated, this case is trivial. If there exists a sequence $\{\omega_n\}\in \#(K)$, then it has $\lim_{n\to\infty} I(\omega_n)= I(\lim_{n\to\infty}\omega_n)=0$ according to the continuity of $I(\omega)$. It indicates $\lim_{n\to\infty}\omega_n\in \#(K)$. Therefore $\#(K)$ is closed.                                    END

From (ii) above it is known that if the function $I(\omega)$ is differentiable, its first order differentiation must be less than the unity. And the result (v) is consistent with Theorem (4.23), which is verified in another way again.

The independent radius provides us a relatively simple technique investigating complete points by the notion of local linear independence. Actually the independent radius is analogous with completeness radius [23].

*Reproducing Kernels*

The function spaces we treated so far belong to $L^2$. The special properties of the functions considered, for example, the bounded property, will supply additional information about the functions, in which we call the reproducing kernel Hilbert space. A reproducing kernel Hilbert space is a Hilbert space of functions in which pointwise evaluation is a continuous linear functional. To be precise, let $H_1$ be a Hilbert space whose elements are real or complex-valued functions defined on a set $J\subseteq \mathbb{R}$. We say that $H_1$ is a reproducing kernel Hilbert space if every linear map of the form $K_\omega: f \to f(\omega)$ from $H_1$ to the complex numbers is continuous (bounded) for any $\omega$ in $J$. By the Riesz representation theorem, every bounded linear functional on $H_1$ arising from an inner product, there is an element $K_\omega \in H_1$ such that $f(\omega) = <f, K_\omega>$ for every $f \in H_1$. The kernel function $K(\omega, t)$ on $J \times J$ ($I=J$) is called the reproducing kernel of $H_1$, defined by $K(\omega, t) = <K_t, K_\omega>$.

The integral transform by the reproducing kernel induces that $\tilde{f}(\omega) = <f, K_\omega> = f$. The complete points can be "seen" from gaps of $f(t)$ in $H_1$. It shows that complete points of the reproducing kernel can always be described explicitly by the investigation of supports of functions in their defined space. But in fact the reproducing kernel space is often defined by a mapping space and we cannot readily "see" the gaps of functions.

Example 4.53. The bandlimited space $B_\Omega = \{ f \in L^2(\mathbb{R}) : \text{support } \hat{f} \subset [-\Omega, \Omega]\}$ is a closed subspace of $L^2(\mathbb{R})$, only defined through the functions in frequency domain conveniently. For $f$ in $B_\Omega$ we have

$$f(\omega) = \int_{\mathbb{R}} f(t) \frac{\sin \Omega(\omega - t)}{\pi(\omega - t)} dt \tag{4.34}$$

The functions that are subject to the formula (4.34) constitute $B_\Omega$. The basis function is $\{\frac{\sin \Omega(t - t_i)}{\pi(t - t_i)}\}$, $t_i = i\pi/\Omega$. Their linear combinations are linearly independent on any interval for the analytic property of functions in $B_\Omega$. That is, any point $\omega \in \mathbb{R}$ is complete.

*Translation Invariant Kernels*

A translation invariant space provides a good representation for signals with the pattern of translation invariance, which is of importance in construction of wavelets and has many applications throughout mathematics and engineering. Usually $T_\alpha$ is adopted to denote unitary operator $T_\alpha \varphi(t) = \varphi(t-\alpha)$, called the translation operator and often suppose integer translations $\alpha = n$ ($n \in \mathbb{Z}$). A translation invariant space $H_1 \subset L^2(\mathbb{R})$ is a space of functions $\{\varphi(t-n)\}$ that is invariant under $T_n$ ($n \in \mathbb{Z}$). There is a result concerning linear independence of $\{\varphi(t-n)\}$.

Proposition 4.54. Let a nonzero function $\varphi \in L^2$ bandlimited on $[-\pi, \pi]$, then $\{\varphi(t-n)\}_{n \in \mathbb{Z}}$ are linearly independent on the real axis.

Proof: Suppose there is a numerical vector $(a_1, a_2, \ldots) \in l^2$, such that

$$\sum_{n=-\infty}^{\infty} a_n \varphi(t-n) = 0 \tag{4.35}$$

Take the Fourier transform on both sides of (4.35), and there is

$$(\sum_{n=-\infty}^{\infty} a_n e^{-in\omega}) \hat{\varphi}(\omega) = 0 \tag{4.36}$$

Since $\varphi$ is not zero, its Fourier transform $\hat{\varphi} \neq 0$ on the whole frequency axis. It requires

$$\sum_{n=-\infty}^{\infty} a_n e^{-in\omega} = 0 \tag{4.37}$$

But $\{e^{-in\omega}\}$ is the basis in $L^2[-\pi, \pi]$. It is linearly independent on $[-\pi, \pi]$. Equality (4.35) doesn't hold except $(a_1, a_2, \ldots) = \mathbf{0}$.    END

When the function sequence $\{\varphi(t-n)\}$ changes to a kernel function $\varphi(t-\omega)$ ($\omega \in \mathbb{R}$), $\varphi$ bandlimited on $I_1$ (a bounded interval) assumed, its associated integral transform is

$$\tilde{f}(\omega) = \int_{\mathbb{R}} f(t) \varphi(t-\omega) dt \tag{4.38}$$

The complete set of $\varphi(t-\omega)$ (($\omega, t) \in \mathbb{R}^2$) is $\#(\varphi(t-\omega)) = \#(e^{-i\omega\xi} \hat{\varphi}(\xi))$ from (iv) of

Theorem (4.24). The weighted exponential kernel defines on $((\omega, \xi) \in \mathbb{R} \times I_1)$. As $\infty \in$ #($e^{-i\omega\xi}\hat{\varphi}(\xi)$), then $\infty \in$ #($\varphi(t-\omega)$). Therefore $\tilde{f}(\omega)$ doesn't vanish at infinity.

The above result also can be understood as follows. Transform (4.38) is represented in form of convolution.

$$\tilde{f}(-\omega)=f(\omega)*\varphi(-\omega) \qquad (4.39)$$

Taking Fourier transform on both sides of (4.39), there is

$$F\tilde{f}(-\omega)=\hat{f}(\xi)\hat{\varphi}(-\xi), \qquad (4.40)$$

where $F\tilde{f}(-\omega)$, $\hat{f}(\xi)$ and $\hat{\varphi}(-\xi)$ are Fourier transforms of $\tilde{f}(-\omega)$, $f(\omega)$ and $\varphi(-\omega)$ respectively. $F\tilde{f}(-\omega)$ is with a finite segment because of the bandlimited function $\varphi$ from (4.40). Hence its inverse transform $\tilde{f}(-\omega)$ disperses at infinity and so $\tilde{f}(\omega)$.

*C. Generalizations: assembly complete function point set*

In this part the simple complete point is extended to a set of points which is called an assembly complete function point set (ACFPS) that no frequency gaps cover this set entirely. Suppose a point set $\{\omega_i\}$ is an ACFPS, and its neighborhood $V_r(\{\omega_i\})$ agrees with the union of neighborhoods $V_{r_i}(\omega_i)$ of $\omega_i$ for any $i$. Then the kernel function is $V_r(\{\omega_i\})$-complete. There is a trivial case that when the sequence $\{K(\omega_i, t)\}$ is complete in the defined space $H_1$, it is clear $\{\omega_i\}$ is an ACFPS. For example, the exponential kernel $e^{-i\omega t}$ ($t \in [-\pi, \pi]$) possesses a natural ACFPS $\{n\}$ ($n \in \mathbb{Z}$). When the integral kernel is degenerated, the ACFPS always exists even with some good properties, for example, stable reconstruction, same to a simple complete point. The finite dimensional kernels are easily addressed. However, an ACFPS of an infinite dimensional kernel can be very complicated. The number of elements in an ACFPS may be finite or be infinite. Here the main focus is taken on the behaviors of functions within the neighborhood of an ACFPS since they are like a basis in some properties sometimes. Following the way of addressing stability of nonharmonics in [23], the problem that a set that is close to an ACFPS shares common properties with this ACFPS is to be resolved in this subsection. Some known results with respect to "closeness" will be introduced.

Theorem 4.55. Let $\{\varphi_n\}$ be a basis for an Hilbert space $H$, and suppose $\{\psi_n\}$ is a sequence of elements of $H$ such that

$$\left\|\sum_{i=1}^{n} c_i(\varphi_i - \psi_i)\right\| \leq \lambda \left\|\sum_{i=1}^{n} c_i \varphi_i\right\|$$

for some constant $\lambda$, $0 \leq \lambda < 1$, and all choices of the scalars $c_1$, ..., $c_n$ ($n=1, 2, 3, ...$). Then $\{\psi_n\}$ is a basis for $H$ equivalent to $\{\varphi_n\}$.

This fundamental stability criterion is originally presented about the Banach space. It is due to Paley and Wiener historically the first. This theorem was then extended more immediately by Krein, Milman and Rutman. These two results can be found in [23].

Theorem 4.56. If $\{\varphi_n\}$ is a basis for a Hilbert space $H$, then there exist numbers $\varepsilon_n > 0$ with the following property: if $\{\psi_n\}$ is a sequence of vectors in $H$ for which

$$\|\varphi_n - \psi_n\| < \varepsilon_n, \ n=1, 2, ...,$$

then $\{\psi_n\}$ is a basis for $H$ equivalent to $\{\varphi_n\}$.

The similar and more refined results were obtained and extended to the discrete frame by Christensen [24].

Theorem 4.57. Let $\{\varphi_n\}$ be a frame for a Hilbert space $H$ with frame bounds $A$, $B$ and let $\{\psi_n\}$ be a sequence in $H$. If there exist $\lambda$, $d \geq 0$ such that $\lambda + d/\sqrt{A} < 1$ and

$$\left\|\sum_{i=1}^{n} c_i(\varphi_i - \psi_i)\right\| \leq \lambda \left\|\sum_{i=1}^{n} c_i \varphi_i\right\| + d\left(\sum_{i=1}^{n} |c_i|^2\right)^{1/2}$$

for all finite scalar sequences $\{c_i\}$. Then $\{\psi_n\}$ is a frame for $H$ with bounds $A(1-(\lambda+d/\sqrt{A}))^2$, $B(1+(\lambda+d/\sqrt{B}))^2$. Moreover, if $\{\varphi_n\}$ is a Riesz basis, then $\{\psi_n\}$ is a Riesz basis.

Theorem 4.58. Let $\{\varphi_n\}$ be a frame for $H$ with bounds $A$, $B$, and let $\{\psi_n\}$ be a sequence in $H$. If there exists a constant $\lambda < A$ such that

$$\sum_{n=1}^{\infty} |<f, \varphi_n - \psi_n>|^2 \leq \lambda \|f\|^2, \forall\ f \in H.$$

Then $\{\psi_n\}$ is a frame for $H$ with bounds $A(1-\sqrt{\lambda/A})^2$, $B(1-\sqrt{\lambda/B})^2$.

The stability or perturbation of frames has been studied explicitly and has a precise theory. This theory can be employed to investigate the perturbation of an ACFPS. If the integral kernel is finite dimensional, a point set close to an ACFPS will also be an ACFPS. Given an ACFPS $\{\omega_i\}$, their corresponding functions in $V_r(\{\omega_i\})$ are denoted by $\{K(\xi_j, t)\}$, $\xi_j \in V_r(\{\omega_i\})$. The symbol $r$ stands for $\{r_i\}$. When we say any $r > 0$, it means any choices $r_i > 0$ for all $i$.

Theorem 4.59. Let $K(\omega, t)$ $((\omega, t) \in J \times I)$ be an $N$-dimensional integral kernel, $H_1$ be its defined space and $\{\omega_i\}$ be an ACFPS. Suppose $\{\omega'_i\} \in J$. If for any neighborhood $V_r(\{\omega_i\})$, and any functions $\{K(\xi_j, t)\}$ $(j=1, \ldots, N)$ in $V_r(\{\omega_i\})$, there always exists a sequence $\{K(\xi'_j, t)\}$ $(j=1, \ldots, N)$ in $V_s(\{\omega'_i\})$ such that

$$\left\|\sum_{j=1}^{n} c_j(K(\xi_j, t) - K(\xi'_j, t))\right\| \leq \lambda \left\|\sum_{j=1}^{n} c_j K(\xi_j, t)\right\| \qquad (4.41)$$

for some constant $\lambda$, $0 \leq \lambda < 1$, and all choices of the scalars $c_1, \ldots, c_n$ ($n=1, 2, \ldots, N$),

then $\{\omega'_i\}$ is an ACFPS for $K(\omega, t)$.

Proof: Since the kernel is finite dimensional and $\{\omega_i\}$ is an ACFPS, one can always choose functions $\{K(\xi_j,t)\}$ ($j=1, \ldots, N$) in any neighborhood $V_r(\{\omega_i\})$ that $\{K(\xi_j,t)\}$ becomes a basis of $H_1$. But there is a sequence $\{K(\xi'_j,t)\}$ in $V_s(\{\omega'_i\})$ for any $s>0$ admitting (4.41). $\{K(\xi'_j,t)\}$ ($j=1, \ldots, N$) is also a basis according to Theorem 4.55. Then in any neighborhood of $\{\omega'_i\}$, there is a basis no frequency gaps covering this set. Therefore $\{\omega'_i\}$ is an ACFPS for $K(\omega, t)$.  END

It should be noted the two sets $\{\omega_i\}$ and $\{\omega'_i\}$ probably don't have the same number of elements. Also there is a perturbation result about the ACFPS of finite dimensional space to be given without proof as they can be readily understood from Theorem 4.56 and the above proof.

Theorem 4.60. Let $K(\omega, t)$ ($(\omega, t) \in J \times I$) be an $N$-dimensional integral kernel, $H_1$ be its defined space and $\{\omega_i\}$ be an ACFPS. Suppose $\{\omega'_i\} \in J$. If for any neighborhood $V_r(\{\omega_i\})$, and any functions $\{K(\xi_j,t)\}$ ($j=1, \ldots, N$) in $V_r(\{\omega_i\})$, there always exist a sequence $\{K(\xi'_j,t)\}$ ($j=1, \ldots, N$) in $V_s(\{\omega'_i\})$ and numbers $\varepsilon_j>0$ for which

$$\|K(\xi_j,t) - K(\xi'_j,t)\| < \varepsilon_j, j=1, 2, \ldots, N,$$

then $\{\omega'_i\}$ is an ACFPS for $K(\omega, t)$.

For a finite dimensional kernel, there are routine results similar to Theorem 4.59-4.60 using Theorem 4.57-4.58. When the continuity is imposed on integral kernels, a set very close to an ACFPS also can be an ACFPS by analogy of Kadec's 1/4-theorem on Fourier integral kernel [23].

Proposition 4.61. Suppose $K(\omega, t)$ is strong continuous on $J \times I$, its defined space $H_1$

and $\{\omega_i\}$ is an ACFPS. Then there is a constant $L>0$, $\{\omega'_i\}$ is also an ACFPS if $|\omega'_i - \omega_i|<L$ for each $i$.

Proof: Suppose $\{K(\xi_j,t)\}$ ($j=1, 2, \ldots$) within any neighborhood $V_r(\{\omega_i\})$ is complete in $H_1$. If in any neighborhood $V_s(\{\omega'_i\})$ the sequence $\{K(\xi'_j,t)\}$ ($j=1, 2, \ldots$) is always incomplete for any $L>0$, then there is a nonzero function $f$ such that $<f, K(\xi'_j,t)>=0$ ($j=1, 2, \ldots$). There is

$$|\int_I K(\xi_j,t)f(t)dt|=|\int_I K(\xi_j,t)f(t)dt - \int_I K(\xi'_j,t)f(t)dt|$$
$$=|\int_I (K(\xi_j,t)-K(\xi'_j,t))f(t)dt|\leq C|\Delta\xi_j|\|f\| \ (j=1, 2, \ldots). \quad (4.42)$$

Here $\Delta\xi_j=\xi_j-\xi'_j$ ($j=1, 2, \ldots$), $|\Delta\xi_j|<L+M_r+M_s$, $M_r$, $M_s$ are maximal values of $\{r_i\}$ and $\{s_i\}$ respectively. $C$ is a constant related to the strong continuity of $K(\omega,t)$. This inequality indicates that the value $<f, K(\xi_j, t)>$ tends to zero as $\Delta\xi_j$ becomes infinitesimals if $L$, $r$ and $s$ are arbitrarily small. But $<f, K(\xi_j, t)>$ cannot be zero for all $j$ simultaneously. For some $j$, $<f, K(\xi_j, t)>$ will be greater than $C|\Delta\xi_j|\|f\|$ when $\Delta\xi_j$ is enough small since $\{K(\xi_j, t)\}$ is complete in $H_1$. It implies $f=0$ as (4.42) is satisfied. Then there is $L>0$ such that $\{\omega'\}$ is an ACFPS if $|\omega'_i - \omega_i|<L$ for each $i$.     END

## V. Density, Sampling and QUP

The complete point theory, particularly developed for the uncertainty principle, has been built. It will be shown that the qualitative uncertainty principle in the $L^2$ norm sense can be well handled under this framework. In this section the nature of QUPs for the usual time frequency integral transforms, namely Fourier transform, Wigner, Gabor and wavelet analysis is revealed by identification of their complete points respectively. The QUP is thus investigated based on the complete point theory. Its relations to density of functions, sampling theory are also disclosed.

Let ($K$, $H_1$, $H_2$) be the 3-tuple as defined in (4.1), ($\omega$, $t$)$\in J \times I \subseteq \mathbb{R}^2$. Its inverse 3-tuple is ($K^{-1}$, $H_2$, $H_1$). $K^{-1}$: $H_2 \to H_1$ is the inverse of $K$, whose integral kernel is denoted by $K^{-1}(t, \omega)$. Set arbitrary subsets $I_1 \subset I$ and $J_1 \subset J$ such that $|I_1|<|I|$ and $|J_1|<|J|$. In this section it agrees that when we say $K(\omega, t)$ defined on $J_1 \times I$, it means $K(\omega, t)$ equals zero except this set.

Theorem 5.1. The uncertainty principle (4.1) holds for that

(i) When $J \times I$ is a compact support of $\mathbb{R}^2$, $J$, $I$ are complete sets corresponding to $K(\omega, t)$ defined on any $I_1$ ($t \in I_1$) and $K^{-1}(t, \omega)$ on any $J_1$ ($\omega \in J_1$) respectively.

(ii) When $J \times I = \mathbb{R}^2$, the point $\omega=\infty$ is complete for $K(\omega, t)$ ($t \in I_1$) also $t=\infty$ complete for $K^{-1}(t, \omega)$ ($\omega \in J_1$).

This theorem is a consequential result of the complete theory. It has other versions under conditions where the sets of $J$ and $I$ are slightly altered from (i), (ii). It shows that the uncertainty principle, which demands strict conditions, is not naturally met and it will be violated when these conditions are lost. However, many frequently applied integral transforms either in mathematics or in engineering always agree with the uncertainty principle. We argue that integral transforms subject to the uncertainty principle are all "good" ones, while those transforms where the uncertainty principle is invalid are "bad" for applications in practice with faults. The violation of uncertainty principle will cause unfavorable results, for example, unstable recovery of a timelimited signal, the loss of uniqueness of transforms, moreover, an intuition that the more information is acquired with the longer data segment will conflict. This may be why the uncertainty principle is everywhere in applications of many areas.

Theorem 5.2. For nonzero $f \in L^2(\mathbb{R})$ arbitrary, $f$ and its Fourier transform $\hat{f}$ cannot be both compactly supported.

Proof: The integral kernels associated with the Fourier transform and its inverse are $K(\omega, t)= e^{-i\omega t}$ and $K^{-1}(t, \omega)= K^*(\omega, t)= e^{i\omega t}$ respectively. $K^{-1}(t, \omega)$ $((t, \omega) \in \mathbb{R} \times J_1$, any $J_1 \subset \mathbb{R}$ a compact support) is proved in the preceding section that $t=\infty$ is complete and also the complete point $\omega=\infty$ for $K(\omega, t)$ $((\omega, t) \in \mathbb{R} \times I_1$, any $I_1 \subset \mathbb{R}$ a compact support). From (ii) of Theorem 5.1, it is known that $f$ and its Fourier transform $\hat{f}$ cannot be both compactly supported except $f=0$.     END

Because of the fundamental limitation by uncertainty principle for classical Fourier transform, Wigner distribution was thus introduced with expectation to reduce this limitation in the time-frequency space (also known as phase space in quantum mechanics). The Wigner distribution is

$$Wf(\omega, t)= \int_{-\infty}^{\infty} f(t+\frac{x}{2})f^*(t-\frac{x}{2})e^{-i\omega x}dx . \qquad (5.1)$$

In accordance with the physical significance of Wigner distribution, it should claims that the occupations of time $t$ and frequency $\omega$ of any nonzero $L^2$ function cannot be simultaneously compact if Wigner distribution agrees with the uncertainty principle. But a natural extension of uncertainty principle (4.1) to the multi-index kernel is that $f(x)$ and $Wf(\omega, t)$ aren't both with compact supports. As a matter of fact, the two statements of uncertainty principle are compatible. When one looks into the relation between supports of $f$ and $Wf$, it shows that the support in $t$ of $Wf(\omega, t)$ for all

$\omega$ is included in the support of $f(x)$ [13]. And $f$ is surely supported in a bounded interval if $Wf$ is with compact support in $t$. It is easily seen from the following fact. Suppose $Wf(\omega, t)=0$ when $|t|>t_0$. It means $f(t+\frac{x}{2})f^*(t-\frac{x}{2})$ vanishing for $x\in \mathbb{R}$, $|t|>t_0$. Therefore $f$ must be timelimited. Define $|\Sigma(Wf(\omega, t))|=m((\omega, t):Wf(\omega, t)\neq 0)$ denoting Lebesgue measure of the set constructed by points $(\omega, t)$ in which $Wf(\omega, t)\neq 0$. Therefore we have the uncertainty principle for Wigner distribution as follows.

Theorem 5.3. If $|\Sigma(Wf(\omega, t))|<\infty$, then $f=0$.

Proof: To make $|\Sigma(Wf(\omega, t))|$ bounded, it is required $Wf$ is timelimited and bandlimited both. So is the function $g_t(x)=f(t+\frac{x}{2})f^*(t-\frac{x}{2})$ timelimited in terms of $x$. Wigner distribution then becomes Fourier transform of $g_t(x)$ with respect to the parameter $t$. Obviously $Wf(\omega, t)$ is not bandlimited for any given $t$ except $g_t(x)=0$ or $f(x)=0$. END

Gabor functions[*] are a group of functions obtained from a signal by time and frequency shift. Its continuous kernel function is $\varphi_{\omega,t}(x)=\varphi(x-t)e^{i\omega x}$, $(\omega, t, x)\in \mathbb{R}^3$ translated by the time $t$, modulated by the frequency $\omega$ and $\varphi(x)$ is a real and symmetric function in $L^2(\mathbb{R})$. They give a complete representation of $L^2(\mathbb{R})$ and can recover signals from its transform stably. The Gabor transform is denoted by

$$G_\varphi f(\omega, t)=\int_{-\infty}^{\infty} f(x)\varphi(x-t)e^{-i\omega x}dx \qquad (5.2)$$

and its inverse

$$f(x)=\int_{-\infty}^{\infty}\int_{-\infty}^{\infty} G_\varphi f(\omega,t)\varphi(x-t)e^{i\omega x}d\omega dt. \qquad (5.3)$$

The energy is conserved such that $\|f(x)\|=\|G_\varphi f(\omega, t)\|/2\pi$. Furthermore the metric is preserved that $<G_\varphi f, G_\varphi g>=<f, g>/\|\varphi\|$.

---

[*] In other occasions they refer to Gaussian functions $a\exp(-bt^2)$ especially.

The investigation of complete points about time-frequency integral transforms from $H_1 \subseteq L^2(\mathbb{R})$ to $H_2 \subseteq L^2(\mathbb{R}^2)$ are somewhat slightly different from the transform whose mapped space also belonging to $L^2(\mathbb{R})$. In the single index integral transform the complete point only depends on the kernel function itself. For example, one just set the variable $t$ of $e^{i\omega t}$ defined on a bounded interval $I$ to study complete points hence the corresponding QUP. But the complete point and QUP in time frequency transforms don't be accessed simplify by defining the kernel on a compact support. They should be related to the support of the original signal $f$. In this section it is always supposed to be timelimited. With respect to Gabor transform it is clear the kernel $\varphi_{\omega,t}(x)$ (($\omega$, $t$, $x$) $\in \mathbb{R}^2 \times I, I \subseteq \mathbb{R}$) has complete points ($\omega$, $t$)=($\infty$, $t$) for $t$ admitting $f(x)\varphi(x-t) \neq 0$ on $I$ known from the property of $\{e^{i\omega t}\}$. And when the quantity $\omega$ is fixed, the Gabor transform is then associated with the translation invariant kernel $\varphi(x-t)$. The point ($\omega$, $t$)=($\omega$, $\infty$) for any $\omega$ is also complete if $\varphi$ is bandlimited according to the result about (4.38). In above two cases, supports of a function $f$ and its transform $G_\varphi f$ aren't both bounded. And these two cases cannot be disappearing simultaneously. Hence QUP holds for Gabor transform. On the other hand, the inverse kernel $\varphi_x(\omega, t)$ (($x$, $\omega$, $t$) $\in \mathbb{R} \times J \times I$, $J$ and $I$ are compact) has complete point $x=\infty$. It also obtains the QUP for Gabor transform.

Theorem 5.4. If $|\Sigma(f)||\Sigma(G_\varphi f)| < \infty$, then $f=0$.

The Gabor transform, or short time Fourier transform is often utilized to localize the Fourier frequency in the neighborhood of $x=t$. The magnitude of $G_\varphi f$ measures the energy distribution of $f$ in the time-frequency plane specified by the localization of

$\varphi_{\omega,t}(x)$. The Gabor transform $G_\varphi f$ provides the information of time and frequency both. It is an overcomplete representation of $f$. Actually the function $f$ can be recovered from only its frequency information and even partial information around $\omega=\infty$ when the timelimited $f$ is supposed to be timelimited. By the time location information of $G_\varphi f$ the recovery is also possible given a well behaved window function $\varphi(x)$. But it has been pointed out in literature that it is the redundancy information by the overcompleteness of $\varphi_{\omega,t}(x)$ that makes time-frequency a favorable idea.

The wavelet is another time-frequency tool. It can be achieved by using a scaling factor instead of modulation in Gabor transform. There is a good controllable compromise between the time and frequency concentrations in a wavelet transform. A wavelet kernel is obtained from a function $\varphi \in L^2(\mathbb{R})$ with a zero average by scaling it by $s$ and translating it by $u$ such that $\varphi_{u,s}(t)=\frac{1}{\sqrt{s}}\varphi(\frac{t-u}{s})$, $(u, s) \in \mathbb{R} \times \mathbb{R}^+$. The wavelet transform of $f \in L^2(\mathbb{R})$ at time $u$ and scale $s$ is

$$Wf(u,s)=\int_{-\infty}^{\infty} f(t)\frac{1}{\sqrt{s}}\varphi^*(\frac{t-u}{s})dt. \qquad (5.4)$$

Its reconstruction formula is

$$f(t)=\frac{1}{C_\varphi}\int_0^\infty \int_{-\infty}^\infty Wf(u,s)\frac{1}{\sqrt{s}}\varphi(\frac{t-u}{s})du\frac{ds}{s^2}, \qquad (5.5)$$

where $\varphi$ be a real function such that $C_\varphi=\int_0^\infty |\hat{\varphi}(\omega)|^2/\omega d\omega<\infty$.

Suppose $\varphi$ is bandlimited and the kernel $\varphi_{u,s}(t)$ $((u, s, t)\in \mathbb{R}\times\mathbb{R}^+\times\mathbb{R})$ has complete points $(u, s)=(\infty, s)$ for any $s\in \mathbb{R}^+$ when the signal $f$ is timelimited. This result is evident as $\#(\varphi_{u,s}(t))=\#(\sqrt{s}\hat{\varphi}(s\omega)e^{-iu\omega})$ from property (iv) of Theorem 4.24. Since $\varphi$ is bandlimited and the scale $s$ is a positive value, the points $(u, s)=(\infty, s)$ belong to

#($\sqrt{s}\hat{\varphi}(s\omega)e^{-ius\omega}$). When $\varphi$ is timelimited, $u=\infty$ fails to be a complete point for $\varphi_{u,s}(t)$ as $Wf(u, s)$ will be zero for enough large $u$ when the signal $f$ is timelimited. But one sees the point $s=\infty$ may be the complete point of $\varphi_{u,s}(t)$ for a general selection of mother wavelet $\varphi$. In this case there are some exceptions. For example, when $f$ is odd and $\varphi$ is even, $Wf$ is null for $u=0$ and for any, even an infinite value of $s$. To avoid this failure, the translation $u$ is set great enough. Now suppose QUP doesn't hold for a function $f \in L^2(-T, T)$ and finite values $u_0$ ($u_0$ is large enough with respect to support of $f$) and $s_0$. That is, $Wf(u_0, s)=0$ ($s>s_0$). In general the information associated with $Wf(u, s)$ for $s>s_0$ can be obtained by introducing a scaling function $\phi$ whose modulus of its Fourier transform is defined by

$$|\hat{\phi}(\omega)|^2 = \int_{\omega}^{\infty} |\hat{\varphi}(\xi)|^2 /\xi d\xi \tag{5.6}$$

and the complex phase of $\hat{\phi}(\omega)$ is freely chosen. Set the function $\phi_s(t)=\frac{1}{\sqrt{s}}\phi(\frac{t}{s})$ and $\bar{\phi}_s(t)=\phi_s^*(-t)$. The information $Wf(u, s)(s>s_0)$ can be substituted by $Lf(u, s)$, the low-frequency approximation of $f$ at scale $s$. It has $Lf(u, s)= f * \bar{\phi}_s(u)$. The reconstruction of $f$ can be with the information that corresponds to $Wf(u, s)(s<s_0)$ and $Lf(u, s_0)$ at scale $s_0$ [13]. It has

$$f(t)=\frac{1}{C_\varphi}\int_0^{s_0} Wf(.,s)*\varphi_s(t)\frac{ds}{s^2}+\frac{1}{C_\varphi s_0}Lf(.,s_0)*\phi_{s_0}(t). \tag{5.7}$$

Comparing (5.7) to (5.5), it should be $Lf(., s_0)* \phi_{s_0}(t)=0$ if the QUP is violated. To make this term vanish, its Fourier transform that $F(Lf(., s_0)* \phi_{s_0}(t))=\hat{f}\hat{\bar{\phi}}_{s_0}\hat{\phi}_{s_0}=\hat{f}|\hat{\phi}_{s_0}|^2$, must be zero. It is known $\hat{f}$ disperses at the infinity as $f$ is timelimited. Therefore it requires $\hat{\phi}_{s_0}$ vanish on the whole frequency line. This is in contradiction with (5.6). In this sense we say $(u, s)=(u, \infty)$ (for values of $u$ can be exceptions) is complete in the kernel $\varphi_{u,s}(t)$.

Theorem 5.5. If $|\Sigma(f)||\Sigma(Wf)|<\infty$, then $f=0$.

It is known that when the scale $s$ tends to $\infty$, its resolution is reaching zero; when $s$ tends to 0, the detail of the signal $f$ is observed with an infinitely large resolution. The information corresponding to large scales can be considered that is obtained from lowpass filters. But a timelimited signal cannot be removed only by the lowpass filters, that is, large scale wavelets. The effect of time truncated on wavelet transform $Wf$ will continue to exist at the infinity. The QUP so occurs!

The Wigner distribution, Gabor and wavelet transforms are different ways splitting time-frequency plane, i.e. analyzing frequency of a signal with respect to its time evolution. But these divisions are very restricted in case of more characteristics and fine structures of signals to be resolved. As a generalization, there are multi-index integral transforms, which map the space $L^2(\mathbb{R}^m)$ to $L^2(\mathbb{R}^n)$, adaptable to deal with the data in high dimensions where it is often the realm of soft computing. Now tools in the area of multiscale geometric analysis, for example, ridgelet [144], curvelet [145], grouplet [146], wave atoms [147], etc. can be viewed as belonging to such a multi-index integral transform class. Does QUP also hold for these multi-index integral transforms? QUP will hold for "good" integral transforms as stated in the preceding. For a finite time length signal, it is impossible that its transformed function vanishes on all the indices at their infinities in mapped space. The integral transform is

$$\tilde{f}(\omega_1,\ldots,\omega_n) = \int_I f(t_1,\ldots,t_m) K(\omega_1,\ldots,\omega_n,t_1,\ldots,t_m) dt_1 \cdots dt_m. \qquad (5.8)$$

Set $\boldsymbol{\omega}=(\omega_1,\ldots,\omega_n)$, $\boldsymbol{t}=(t_1,\ldots,t_m)$, $\boldsymbol{t} \in I \subseteq \mathbb{R}^m$, $\boldsymbol{\omega} \in J \subseteq \mathbb{R}^n$ and usually $n \geq m$. The kernel $\boldsymbol{K}(\boldsymbol{\omega},\boldsymbol{t})=K(\omega_1,\ldots,\omega_n,t_1,\ldots,t_m)$ is supposed to be strong continuous inducing an operator $K(\cdot)$ from $L^2(I)$ to $L^2(J)$. Suppose the QUP isn't met, then it has a space $H \subset L^2(I)$ such that $\tilde{f}$ is mapped on a compact support $J_1 \subset J$ for any nonzero function $f \in H$. If $I$ and $J_1$ are both compact supports, then it indicates $K(\cdot)$ a compact operator when the signal is restricted in the subspace $H$. The recovery of $f$ in $H$ from $\tilde{f}$ will become unstable. This is one fault of violation of QUP. Therefore that $\tilde{f}$ doesn't vanish in the infinity in terms of at least one variable (index) $\omega_i$ ensures the satisfaction of QUP This makes the data observation persist in an infinitely long time that is impossible in the real world. A large amount of data is required for an accurate reconstruction in applications. As stated in Section II any "clever" algorithm cannot compensate the loss of information. One would accept a not so accurate reconstruction with a tradeoff between the error and the amount of data. It is interesting noted the good behaved integral operator also becomes compact if the infinite support is truncated to a finite segment. However, the truncated operator can provide good results in numerical analysis.

On some popular integral kernels the uncertainty principle has been investigated. While, on the other hand there are papers concerning uncertainty relations for Fourier series even orthonormal sequences [141], [142], [143]. It says there does not exist an orthonormal basis for $L^2(\mathbb{R})$ that this orthonormal basis and its Fourier transform are sharply localized. In this part it will be shown that the QUP, not restrictive on the orthonormal sequences, but for function sequences can be readily derived from the uncertainty principle of integral kernel by sampling method.

Any signal $f$ in the Hilbert space $H$, the set $\Lambda=\{t_n\}$ is called a sampling set when the signal $f$ can be recovered from its samples $\{f(t_n)\}$, $t_n \in \Lambda$ that $A\|f\|^2 \leq \sum |f(t_n)|^2 \leq B\|f\|^2$ holds with positive constants $A$, $B$. The explicit sampling formula was found that it is associated with the frame in Hilbert spaces, for example, the sinc kernel in Nyquist sampling formula is a basis for the bandlimited space. Let the integral kernel $K(\omega,t)$ be a continuous frame with bounds $B \geq A > 0$ inducing a map from $H_1$ to $H_2$. It is supposed to be sampled to a discrete frame, that is, it has a representation

$$K(\omega,t)=\sum_{n=1}^{\infty}\psi_n(\omega)K(\omega_n,t) \qquad (5.9)$$

Here $\{K(\omega_n, t)\}$ constitutes a discrete frame with bounds $B_1 \geq A_1 > 0$ for $H_1$. Therefore the mapped space $H_2$ can be sampled through $K(\omega,t)$ that

$$\tilde{f}(\omega)=\int_I f(t)K(\omega,t)dt$$
$$=\sum_{n=1}^{\infty}\psi_n(\omega)\int_I f(t)K(\omega_n,t)dt$$
$$=\sum_{n=1}^{\infty}\tilde{f}(\omega_n)\psi_n(\omega) \qquad (5.10)$$

It is easily verified that

$$\frac{A_1}{B}\|\tilde{f}\|^2 \leq A_1\|f\|^2 \leq \sum_{n=1}^{\infty}|\tilde{f}(\omega_n)|^2 = \sum_{n=1}^{\infty}|<f(t),K(\omega_n,t)>|^2 \leq B_1\|f\|^2 \leq \frac{B_1}{A}\|\tilde{f}\|^2. \qquad (5.11)$$

Hence $\Lambda=\{\omega_n\}$ is a sampling set for any signal $\tilde{f}(\omega)$ in $H_2$. If $K(\omega,t)$ is symmetric or $\{K(\omega, t_n)\}$ constructs a discrete frame in $H_2$, any signal $f(t)$ of $H_1$ also can be sampled through samples $\{f(t_n)\}$.

The problem that whether a continuous frame is sampled to a discrete frame also remains open. But for some usual continuous frames which are the kernels of integral transforms in uncertainty principle, there are celebrated results relating continuous frames to discrete frames by density of sequences [135-140]. The uncertainty principle for function sequences can be derived from uncertainty principle of integral transforms with sampling technique.

Let $\Lambda \subset \mathbb{R}$ be a set without accumulation points and the sequence is called separated if $\inf_{\omega_i,\omega_j \in \Lambda, i \neq j} |\omega_i - \omega_j| > 0$. The upper and lower uniform Beurling densities are defined as:

$$D^+(\Lambda) = \lim_{r \to \infty} \frac{n^+(r)}{r}$$

and
$$D^-(\Lambda) = \lim_{r \to \infty} \frac{n^-(r)}{r}$$

where $n^+(r)$ and $n^-(r)$ denote counting functions respectively the largest and smallest number of points of $\Lambda$ in an interval of length $r$. If these two densities coincide, $D(\Lambda) = D^+(\Lambda) = D^-(\Lambda)$ is used to take the uniform density of $\Lambda$.

With respect to the exponential system $E(\Lambda) = \{\exp(i\omega t), \omega \in \Lambda, t \in I \subset \mathbb{R}\}$, $I$ a single interval, there is a classical result. If $D^-(\Lambda) > |I|/2\pi$, then $E(\Lambda)$ is a frame in $L^2(I)$ and if $D^-(\Lambda) < |I|/2\pi$, then $E(\Lambda)$ is not a frame in $L^2(I)$. From the complete point theory, the functions within neighborhoods of complete points can be complete in their space. However, the functions around a regular complete point construct an unseparated sequence which is not stable proved Section IV. Then the infinity becomes the only choice for the complete point around which their functions provide a frame. The density is such a notion bridging function sequences and complete points. If $D^-(\Lambda) > |I|/2\pi$, then $\hat{f}(\omega) = \sum_{\omega_n \in \Lambda} \hat{f}(\omega_n) \psi_n(\omega)$. Note that $\{\psi_n(\omega)\}$ is associated with the exponential system $E(\Lambda)$. Although $\hat{f}(\omega)$ disperses to the infinity in the above, $\{\hat{f}(\omega_n)\}$ is not certain to persist at the infinity. For example, when the rectangle window $f = \chi_I$ is analyzed in the system $E(\Lambda) = \{\exp(int), n \in \mathbb{Z}, I=(-\pi, \pi)\}$, all the elements of $\{\hat{f}(\omega_n)\}$ are zeros except $\hat{f}(\omega_0)$.

Now set $\{\psi_n(\omega)\}$ a frame in $L^2(-\pi, \pi)$, $I_1 \subset (-\pi, \pi)$ and $|I_1| < 2\pi$, $f(t)$ defines on $(-\pi, \pi)$ but vanishes on $(-\pi, \pi)\backslash I_1$. It has $f(t) = \sum_{n \in \mathbb{Z}} \hat{f}(n) e^{int}$. But we see $\{\hat{f}(n)\}$ cannot vanish at the infinity, otherwise the linear combination of finite exponential functions will be zero in the set $(-\pi, \pi)\backslash I_1$, which is impossible as finite number of exponential functions are linear independent on any set of Lebesgue measure nonzero. The linear independence of finite number of exponential functions can be readily verified from their constituted Wronskian determinant, which is a Vandermonde determinant. This is a version of uncertainty principle for Fourier series.

Theorem 5.6. Suppose $\{e^{int}, n \in \mathbb{Z}\}$ defines on $(-\pi, \pi)$ and a function $f(t)$ defines on the set $I_1$. If $|I_1| < 2\pi$, then $\{\hat{f}(n)\}$ disperses at the infinity.

Also it has $\hat{f}(\omega) = \sum_{n \in \mathbb{Z}} \hat{f}(n) \psi_n(\omega)$. One knows that the density of zero points of $\hat{f}(\omega)$ cannot be greater than 1 from the above theorem. More extensively there is the following corollary.

Corollary 5.7. The density of zeros of a function bandlimited within a single interval $I$ cannot be greater than $|I|/2\pi$.

If the kernel $K(\omega, t)$ can be sampled to a discrete frame $\{K(\omega_i, t)\}$, the satisfaction of uncertainty principle similar to Theorem 5.6 depends on the linear independence of any finite number of elements in $\{K(\omega_i,t)\}$, which is called finitely independent or simply independent [112]. Obviously if $K_\omega(t)$ is analytic in terms of $t$ and $H_2$ is a sampling space, then a proposition similar to Theorem 5.6 holds for $\{K(\omega_i,t)\}$. Reversely if the QUP of a function sequence holds, then $\{\hat{f}(\omega_n)\}$ has infinite number of nonzero elements therefore $\hat{f}(\omega)$ disperses at infinity leading to the QUP for its associated integral kernel.

The *density* is a strong notion addressing redundancies of function sequences. Moreover, it implies the QUP. Suppose there is a positive value $d(|I|)$ in terms of the length of a single interval $I$. If the sequence $\{K(\omega,t), \omega \in \Lambda, t \in I \subset \mathbb{R}\}$ constructs a frame of $L^2(I)$ for any compact $I$ when $D^-(\Lambda) > d(|I|)$, then $\omega=\infty$ is a complete point. Otherwise there is a nonzero function $f \in L^2(I)$ and a value $r>0$ such that $<f, K(\omega,t)>=0$ $(\omega \in V_r(\infty))$. Set $\Lambda_1=\{\omega: \omega \in \Lambda \cap V_r(\infty)\}$. Then $\{K(\omega,t), \omega \in \Lambda_1, t \in I \subset \mathbb{R}\}$ is also a frame of $L^2(I)$ as $D^-(\Lambda_1)=D^-(\Lambda)$, where $<f, K(\omega,t)>$ aren't zeros for all $\omega \in \Lambda_1$. Hence it contradicts with the vanishment of $<f, K(\omega,t)>$ $(\omega \in V_r(\infty))$.

## VI. Harmonic Analysis beyond Uncertainty Principle

It has been discovered that the uncertainty principle isn't necessary for all the linear integral transforms, but the uncertainty principle indeed holds for well behaved linear integral transforms. In this section we focus on how to achieve a super-resolution of function decomposition beyond the uncertainty principle. The maximal capacity of linear integral operators on harmonic analysis exploiting the uncertainty principle is to be studied. A general concept of super-resolution is developed to encompass some recent techniques beyond the uncertainty principle providing a unified viewpoint.

Despite the QUP holds in many cases, one may still be able to construct a linear integral transform violating the QUP. The identification of elements in exponential sums is the core task in signal analysis. It is expected that there is an integral transform mapping exponentials to functions in frequency domain with compact supports simultaneously. Such an integral operator would replace the Fourier transform to provide a concise representation for a signal. It would be very useful in analysis of truncated signals with high precision, compressed data transmission and many others. The QUP might be fully violated. Unfortunately the following theorem disproves it.

Theorem 6.1. Let $K(\omega, t)$ be an integral kernel that $K_\omega(t) \in L^2$. If its related integral operator maps a complex exponential function $\{e^{i\lambda t}\}$ ($t \in I$ and $I$ is compact) to a function $\tilde{f}_\lambda(\omega)$ also with compact support for any $\lambda \in \mathbb{R}$, then there is a compact support $J$ such that $K(\omega, t)=0$ ($(\omega, t) \in \mathbb{R} \setminus J \times I$).

Proof: The integral transform is

$$\tilde{f}_\lambda(\omega) = \int_I K(\omega,t) e^{i\lambda t} dt . \qquad (6.1)$$

The function $\tilde{f}_\lambda(\omega)$ depending on the frequency parameter $\lambda$ is also denoted by $\tilde{f}(\lambda,\omega)$. Set $\tilde{f}(\lambda,\omega) = \hat{f}_\omega(\lambda)$, which is the truncated Fourier transform of $K_\omega(t)$ when $\omega$ is fixed and $\lambda$ changes over the real axis in (6.1). The function $\hat{f}_\omega(\lambda)$ spreads to the infinity in terms of $\lambda$. Suppose $\tilde{f}_\lambda(\omega)$ for any $\lambda \in \mathbb{R}$ is restricted within a compact support. It implies that the union of supports of $\{\tilde{f}(\lambda_n,\omega)\}$ transformed from the basis $\{e^{i\lambda_n t}\}$ of $L^2(I)$ will be contained in a bounded set $J$. If such a set $J$ doesn't exist, then the length of union of supports of $\{\tilde{f}(\lambda_n,\omega)\}$ will be infinite. But it is known any function $e^{i\lambda t}$ on $I$ can be expanded by $\{e^{i\lambda_n t}\}$ therefore the defined interval of $\tilde{f}_\lambda(\omega)$ will be infinity. It is a violation of our supposition. Therefore the support of $\tilde{f}_\lambda(\omega)$ is contained in $J$ for any $\lambda \in \mathbb{R}$. It suggests $\hat{f}_\omega(\lambda)=0$ ($\lambda \in \mathbb{R}$, $\omega \in \mathbb{R} \setminus J$) which is transformed from functions $K_\omega(t)$ ($t \in I$, $\omega \in \mathbb{R} \setminus J$). It should be $K_\omega(t)=0$ ($\omega \in \mathbb{R} \setminus J$). END

The integral kernel $K(\omega, t)$ is expected to mirror $\{e^{i\lambda t}\}$ in the frequency domain and replaces the Fourier frequency with "$K$-frequency" on analyzing the spectrum of a time signal $f \in L^2(I)$. But Theorem 6.1 shows that a large portion of "$K$-frequency" information of the spectrum will be lost since $K_\omega(t)$ vanishes outside a bounded set. The transform (6.1) hence doesn't provide a complete representation of the time signal in "$K$-frequency" domain. Because of the function system $\{e^{i\lambda t}\}$ actually spans $L^2(I)$, this theorem also says it cannot maps any function of $L^2(I)$ to the function in "$K$-frequency" domain compactly supported. There is a corollary equivalent to the above theorem.

Corollary 6.2. Let $I$ be a single compact interval and $\Lambda$ be a discrete set. If an integral operator maps $\{\exp(i\omega t),\ \omega \in \Lambda \subset \mathbb{R},\ t \in I,\ D^-(\Lambda) > |I|/2\pi\}$ to $\tilde{f}_\lambda(\omega)$ with a compact support, then there is a bounded set $J$ such that $K(\omega, t)=0$ (($\omega, t) \in \mathbb{R} \setminus J \times I$).

If a function sequence $\{\varphi_i(t)\}$ is complete in $L^2(I)$, then they also cannot be mapped to functions with compact supports all.

Corollary 6.3. If a function sequence $\{\varphi_i(t)\}$ is complete in $L^2(I)$, $\tilde{f}_i(\omega)=K\varphi_i(t)$ with a compact support for all *i*, then there is a bounded set *J* such that $K(\omega, t)=0$ $((\omega, t) \in \mathbb{R} \setminus J \times I)$.

In summary, The conditions underlying propositions 6.1-6.3 which doesn't lead to the full violation of QUP are that their corresponding integral operators are linear and complete in $L^2(I)$. It has been known so far there are several sufficient conditions for the 3-tuple ($H_1$, $H_2$, $K$) leading to QUP. However, its necessity remains unknown. Here the necessary conditions are partly obtained. If the conditions are removed, then a super resolution can be achieved. There are two ways to bypass the uncertainty principle. Firstly it is about the linearity of integral operators. The uncertainty relations have often taken place as an accompanying phenomenon in linear transforms. If the linearity is rejected and the nonlinearity is introduced, then the uncertainty principle doesn't hold again and even be violated entirely. There are some nonlinear techniques recently presented to get a super-resolution, for example, the nonlinear Fourier transform and generalized Fourier transform [132-134]. These transforms provide a mathematical unification for PDEs and integral representations. The nonlinear method is a large and complicated area to be addressed. The second way bypassing the uncertainty principle keeps the linearity of operators, therefore is comparatively easy to deal with. This way exploits the incompleteness, even its sparsity of an operator. As is well known, $L^2$ is a classical space modeling the vast majority of signals that we encounter in the real world. But for many purposes, for instance, signal extraction or signal decomposition, it is really large in these respects. One reason is that some notions in harmonic analysis will become obscure because of the completeness of integral representation in $L^2$. For example, given a signal constructed by Fourier series with infinite terms, the accurate signal decomposition is to identify each component in this Fourier series. But we know this signal can also be constituted by nonharmonics if their density is large enough. Hence the decomposition is not unique and precise harmonic component extraction becomes impossible theoretically! The second reason is that any computations and numerical analysis can never be realized in $L^2$ actually. $L^2$ is an ideal mathematical object. It is being in the theory, not in the practice. For example, FFT is implemented replacing the Fourier transform. Thus it is needed to work in the subspaces of $L^2$, even finite dimensional spaces. These spaces should be "sparse" in $L^2$ in some sense. For a sparse set of exponentials or other types of signals, there is the corresponding operator mapping them all with compact supports. When the signal is discretized in the finite dimensional space, the assertion like (3.1) isn't yet satisfied. The maximal sparsity of transformed signal is then pursued [27], [36]. So we must discover the structure of subspace in $L^2$ and design an operator specially applied to this subspace.

It has been known the completeness in $L^2$ must be removed in order to make the signal analysis with a super-resolution. But by this method it implies there are signals which cannot be represented by the operator incomplete and noninvertible in $L^2$ [155]. Therefore one needs a cluster of operators which is called an operator family.

Definition 6.4. The union $K = \bigcup_{\lambda \in \Lambda} K_\lambda$ is said to be an operator family of $H$, if $K_\lambda$ ($\lambda \in \Lambda$, $\Lambda$ is an index set) is an operator of Hilbert space $H$.

The index set $\Lambda$ can be discrete or continuous. Its cardinality decides the complexity of an operator family and computations thereafter. In signal area the dictionary of atoms is such an example. The more diversity and number of atoms in a dictionary, a signal can be represented more accurately. But the computational burden of sparsity pursuit will be increasing and finally will be totally unacceptable. Definition 6.4 comes from a consideration as follows: a "good" (QUP holds) single operator is not suitable to work in the whole $L^2$ space with high resolutions. It only achieves an average performance, not optimal on analyzing signals in $L^2$. Therefore it is required of an operator family. Each operator in this family is optimal for only the certain group of signals. When these operators are collected, it is expected the optimal result is obtained for all the signals in $L^2$.

Definition 6.5. An operator family $K = \bigcup_{\lambda \in \Lambda} K_\lambda$ is said to has a super-resolution if $f \in L^2$ arbitrary, there always exists an operator $K_\lambda$ decomposing this signal under a given rule.

The definition of super-resolution operator family is very general to contain various methods. Given an operator, how one select an operator decomposing a signal well? It depends on given rules. The parametric spectrum estimation is a model-based harmonic analysis. Its parameters are estimated via some optimization methods where operators of a family correspond to these parameters. When the parameters are determined by an optimization method, its associated operator $K_\lambda$ is found. The sparsity is also "a given rule" to select an optimal operator. Its concrete algorithms are $l^1$ optimization and greedy pursuit. As to the ridgelet, curvelet, etc., they are constructed to give a sparse representation for some special signal sets.

The decomposition of finite exponential sums or finite sums of atoms is attractive in signal area. It is unique and suitable in computations. Let $\Sigma_m$ be the collection of all elements which are expressed as a linear combination of $m$ elements of $\{e^{i\lambda t}\}_{\lambda\in\mathbb{R}}$:

$$\Sigma_m = \{f = \sum_{i=1}^{m} a_i e^{i\lambda_i t} : a_i \in \mathbb{R}, \lambda_i \in \mathbb{R}\} \tag{6.2}$$

The space $\Sigma_m$ is not linear: the sum of two elements from $\Sigma_m$ is generally not in $\Sigma_m$, it is in $\Sigma_{2m}$. If $m$ is allowed to be any finite value, then (6.2) is actually a representation with Hamel basis. From a standard result in functional analysis, the cardinality of Hamel basis is uncountable to give a complete representation of $L^2(I)$. If an operator family $K$ is adopted to estimate signals in $\Sigma_m$, the cardinality of $\Lambda$ must be uncountable as the rank of each operator $K_\lambda$ is finite in order to promise optimal results in a finite dimensional space.

In the end an example is presented to describe our scheme.

Example 6.6. A signal is composed of an exponential function and a chirp $f(t)=\exp(i\alpha t) + \exp(i\beta t^2)$ on the real axis. The function $\exp(i\alpha t)$ is well identified by Fourier transform and the chirp can be also estimated by chirplet transform respectively on processing $f(t)$. But Fourier transform and chirplet transform cannot individually resolve two components of $f$. It needs an operator family containing Fourier and chirplet transform both.

In the preparation of this paper, the author noticed recently there were two papers [153], [154] addressing signal representation which can be understood in the author's way.

VII. Conclusions

This paper is about the mathematical investigation of the essence of uncertainty principle, especially for its qualitative version. It is found the QUP is associated with the density of functions. As an extension of the unseparated sequence, the complete point theory is established in attempt to deal with the density of continuous index function systems, which is developed in $L^2$ space. In the $L^p$ ($p\geq 1$, $p\neq 2$) space, the orthogonality is lost and some new techniques are needed to introduce complete points into these spaces. Restricted within linear integral operators and Hilbert spaces, it is discovered that the satisfaction of usual QUP depends on the existence of the complete point $\infty$. The investigation of full violation of QUP implies the termination of linear integral operators which are complete in $L^2$. And it indicates the nonlinear and sparse philosophies.

## Acknowledgement

This paper is devoted to the author's late grandmother, whom he deeply loves and from whom great energy, noble soul but plain life he benefits.## Reference